# TPAD: Hardware Trojan Prevention and Detection for Trusted Integrated Circuits

Tony F. Wu, *Student Member*, Karthik Ganesan, *Student Member*, Yunqing Alexander Hu, *Student Member*, H.-S. Philip Wong, *Fellow*, Simon Wong, *Fellow*, Subhasish Mitra, *Fellow*

***Abstract*— There are increasing concerns about possible malicious modifications of integrated circuits (ICs) used in critical applications. Such attacks are often referred to as hardware Trojans. While many techniques focus on hardware Trojan detection during IC testing, it is still possible for attacks to go undetected. Using a combination of new design techniques and new memory technologies, we present a new approach that detects a wide variety of hardware Trojans during IC testing and also during system operation in the field. Our approach can also prevent a wide variety of attacks during synthesis, place-and-route, and fabrication of ICs. It can be applied to any digital system, and can be tuned for both traditional and split-manufacturing methods. We demonstrate its applicability for both ASICs and FPGAs. Using fabricated test chips with Trojan emulation capabilities and also using simulations, we demonstrate: 1. The area and power costs of our approach can range between 7.4-165% and 0.07-60%, respectively, depending on the design and the attacks targeted; 2. The speed impact can be minimal (close to 0%); 3. Our approach can detect 99.998% of Trojans (emulated using test chips) that do not require detailed knowledge of the design being attacked; 4. Our approach can prevent 99.98% of specific attacks (simulated) that utilize detailed knowledge of the design being attacked (e.g., through reverse-engineering). 5. Our approach never produces any false positives, i.e., it does not report attacks when the IC operates correctly.**

***Index Terms* — Hardware Security, Hardware Trojan, 3D Integration, Resistive RAM, Concurrent Error Detection, Randomized Codes, Split-manufacturing, Reliable Computing.**

## I. Introduction

There is growing concern about the trustworthiness of integrated circuits (*ICs*). If an untrusted party fabricates an IC, there is potential for an adversary to insert a malicious *hardware Trojan* [1], an unauthorized modification of the IC resulting in incorrect functionality and/or sensitive data being exposed [2]. These include (but are not limited to) [3]:

a) Modification of functional behavior through logic changes: For example, extra logic gates may be inserted to force an incorrect logic value on a wire at some (arbitrary) point in time (Fig. 1a) [2].

b) Electrical modification: For example, extra capacitive loading may be placed on a circuit path to alter timing characteristics of the IC (Fig. 1b) [2].

c) Reliability degradation: For example, aging of transistors (e.g., Negative Bias Temperature Instability (NBTI)) may be accelerated, degrading reliability (Fig. 1c) [4].

A hardware Trojan is *detected* when we report malicious modifications to an IC. A hardware Trojan is *prevented* when we stop a hardware Trojan from being inserted. Part of the challenge in detecting or preventing hardware Trojans arises from the fact that the methods for inserting Trojan attacks are numerous (e.g., [2][5] for a comprehensive list).

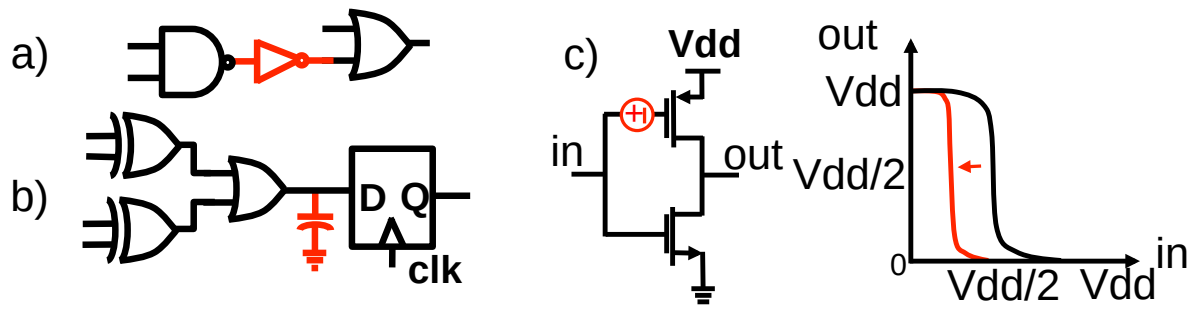

Fig. 1. Examples of hardware Trojans. a) Modification of functional behavior through logic changes; b) Electrical changes such as changing timing characteristics; c) Reliability changes such as accelerating aging with NBTI.

Trojan insertion methods include (but are not limited to):

- Modification of RTL or layout by malicious CAD tools without the designer's knowledge [6].
- IC modification anytime between tapeout and fabrication, since it is possible to reverse engineer (complete or partial) netlists from layouts [7][8]. Such modifications may be applied to all or a sample of fabricated ICs. IC modification can span a wide range from circuit modification to modification of the doping profile of a few transistors [9].
- Malicious reprogramming of systems implemented in reconfigurable platforms (such as FPGAs) with incorrect functionality before startup or during field operation [10].
- Leakage of information, e.g., through fluctuations in IC power consumption [11].

Existing techniques for hardware Trojan detection exhibit one or more of the following limitations (in-depth discussion in Section VII):

1. Some techniques rely on IC testing for Trojan detection, which alone limits the scope of Trojans. For example, Trojans activated by a "time-bomb" [2][12] or accelerated circuit aging [4] may not be detected.
2. Nondestructive visual inspections can be circumvented by carefully hiding Trojans [9]. For example, an adversary can change the doping profile of transistors. Without IC delayering, detection becomes increasingly difficult due to limited imaging depth, especially for 3D ICs [53].
3. Destructive IC testing is effective only when the percentage of ICs attacked is high (e.g., many chips on a wafer must be attacked), as shown in Section VII.A.
4. IC fingerprinting, such as circuit path delays, leakage power, heat or switching activity, relies on statistical models to distinguish between malicious activities vs. variability during manufacturing or normal operation. This makes sufficiently small Trojans undetectable [13]. Moreover, such techniques are prone to reporting false positives [90].
5. Error-detecting codes [16] can be compromised if the code construction is known.

Manuscript received January 9, 2015; revised May 8, 2015; accepted July 20, 2015. Work supported by IARPA Trusted Integrated Chips (TIC) Program.
All authors are with Dept. of Electrical Engineering, Stanford University, Stanford, CA 94305 USA (e-mail: tonyfwu@stanford.edu). S. Mitra is also with Dept. of Computer Science, Stanford University.

This paper presents a Trojan Prevention and Detection (TPAD) architecture for digital systems. TPAD enables test-time and runtime detection of Trojans inserted during fabrication. TPAD prevents insertion of Trojans during fabrication based on sophisticated reverse-engineering of the design being fabricated. TPAD also prevents Trojan insertion during logic synthesis and layout design.

In this paper, we assume that logic synthesis tools, physical design tools, as well as IC fabrication are untrusted, meaning CAD tools and/or foundries can insert Trojans (Fig. 2a). However, we do require the RTL or system specification as well as system assembly to be trusted, meaning that an adversary cannot insert a Trojan during these stages.

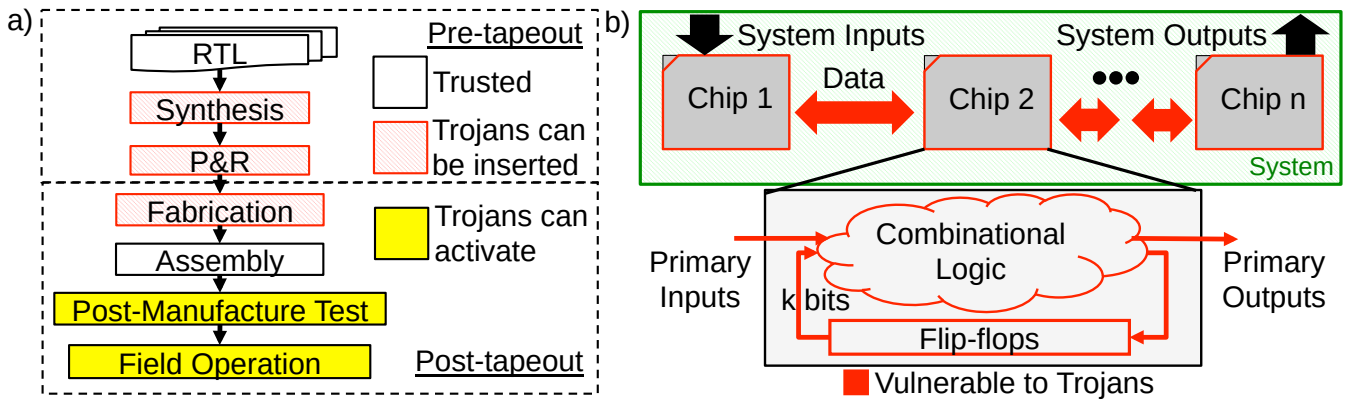

Fig. 2. a) Assumptions on when Trojans can be inserted or activated. b) Assumptions on vulnerable system and chip components.

The following are some of the important TPAD features:

1. TPAD detects hardware Trojans that produce incorrect Boolean logic value(s) that propagate to sequential elements (flip-flops, latches), on-chip memories, or I/O pins.
2. TPAD has a 0% false positive rate, i.e., it does not report Trojan detection when the IC operates correctly.
3. TPAD can be used for Trojan detection concurrently during post-manufacture IC testing and also during field operation (concurrently during system operation or through periodic testing). As a result, Trojans triggered by rare events such as time bomb attacks can be detected. If used in conjunction with system-level recovery techniques (such as those used for fault-tolerant computing), it may be possible to recover from many attacks (Appendix B). Since TPAD significantly improves the observability of signals inside an IC, test-time Trojan detection is expected to be significantly enhanced compared to traditional functional testing.
4. The hardware overhead associated with TPAD is flexible, and can be adjusted based on targeted attack models and implemented functionality.
5. TPAD can be used in conjunction with (or without) split-manufacturing [17]. TPAD can be combined with emerging non-volatile memories (e.g., Resistive RAM or RRAM) and their monolithic 3D integration [18] to reduce hardware overheads.
6. TPAD can be combined with a secure EDA tool flow (Section VI) to prevent insertion of Trojans during logic synthesis and physical design.
7. Currently, TPAD does not target Trojans that leak information without modifying logic values transmitted through the I/Os of the attacked IC, e.g., using a radio inserted maliciously or through fluctuations in power consumption [11].

Our overall approach was briefly outlined without many details in [102]. In this paper, we present an in-depth description of TPAD, experimental results from test chips, as well as various implementation trade-offs with respect to area, power, and delay. The TPAD architecture is introduced in Section II and Section III. Section IV presents an implementation of the TPAD architecture that can be used for any arbitrary digital system. For general processor designs, such as OpenRISC [39], the key results of TPAD are:

1. Area overheads range from 114% to 165% with 34% to 61% power overheads.
2. Near 0% speed impact can be achieved (through additional pipelining in the TPAD architecture).
3. The corresponding detection rate for *uninformed attacks* (attacks that **do not** require detailed knowledge of the design before inserting a Trojan) ranges from 87.5%-99.6%.
4. TPAD prevents sophisticated attacks requiring some detailed knowledge of the implemented design (launched by CAD tools or foundries) with a rate ranging from 96.2% to 99.98% (Section III).

In Section IV.C, we also demonstrate TPAD for fabricated FPGA test chips. Section V presents TPAD implementations for specialized (accelerator) designs, such as a Lempel-Ziv (LZ77) data compressor test chip [45] and a Fast Fourier Transform (FFT) engine [41]. For such accelerators, the key results are:

1. TPAD area overheads can be 7.4%.
2. TPAD detects 99.998% of uninformed attacks.

We further show that for LZ77 data compression, performance metrics such as compression ratio (CR) can be traded off for area overhead with no impact on Trojan detection and prevention.

In Section VI, we discuss how to TPAD can be used to prevent logic synthesis and physical design tools from inserting Trojans. Section VII compares and contrasts existing Trojan detection and prevention approaches vs. TPAD. Section VIII concludes the paper.

Appendix A presents a detailed discussion of various attack types. Appendix B discusses possible approaches to recover from Trojans.

## II. System Architecture Overview

TPAD is derived from the concept of Concurrent Error Detection (*CED*) [19] for fault-tolerant computing. The "classical" CED approach is illustrated in Fig 3a: given a function, an Output Characteristic Predictor (*OCP*) predicts an output characteristic (e.g., parity of output bits) for each input to that function. Another function (*checker*) calculates the actual output characteristic and checks it against the predicted output characteristic. The checker reports an error whenever the two do not match. However, the problem of detecting attacks by hardware Trojans is different than CED for fault-tolerant computing for the following reasons:

1. CED for fault-tolerant computing generally targets single faults (that may occur at random depending on the fault model). For example, CED techniques that predict output parity (or the count of 0's or 1's in an output word) can be easily compromised if an attacker inserts a hardware Trojan which flips output bits of the function such that the number of 1's in a given output word is preserved.

2. An attacker must not be able to derive the OCP; otherwise, the attacker could modify both the function outputs and the OCP outputs such that the checker does not detect errors.
3. CED techniques for fault-tolerant computing generally assume that only one of the units (function, OCP, checkers) is faulty (at any given time). For hardware Trojans, one cannot make such assumptions.

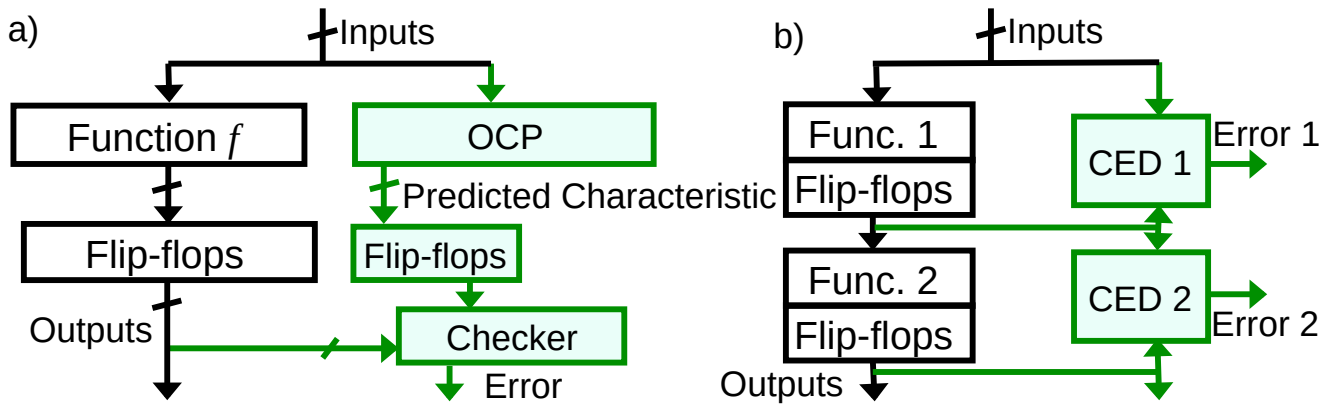

Fig. 3. a) General block diagram of a concurrent error detection scheme; b) Separate CED techniques for separate functions of a chip.

One option is to implement a chip entirely on a reconfigurable platform such as an FPGA (with concurrent error detection for the mapped logic). This can potentially mitigate vulnerabilities to sophisticated attacks that rely on reverse-engineering. However, CMOS FPGAs can incur significant area and performance overheads compared to ASICs (17-27x area and 18-26x performance in [20]). While monolithic 3D integration has been shown to somewhat reduce these costs (5-8x area overheads and 10-15x performance overheads in [21]), these overheads may still be too high for many applications. Further, current trends indicate that an increasing number of accelerators (e.g., signal processing blocks) [22] are being embedded in FPGA chips. These accelerators, along with I/Os, are vulnerable to Trojans. Thus, FPGAs that are manufactured by untrusted foundries can still be vulnerable.

TPAD overcomes the limitations of classical CED techniques. At the same time, it avoids the high overheads of FPGAs through selective hardware programmability (Section III) in Trojan checking circuitry.

Fig. 2b shows a digital system with circuit components vulnerable to Trojan attacks highlighted in red. The system operates in a trusted environment; however, all chips are vulnerable to Trojans. While wires (or channels) between chips may not vulnerable to attacks (since the system may be assembled in a trusted environment), any chip with a Trojan may use them to send incorrect data.

A block diagram of the system architecture with TPAD is shown in Fig. 4. As mentioned in Section I, we assume the system is assembled in a trusted environment. Thus, any Trojan attack within the system will originate from at least one chip. Each chip in the system encodes its outputs and receives encoded inputs. Specifically, Chip 1 outputs data and corresponding check bits so Chip 2 can use them to verify the data (Section II.A-B). Encoded error signals sent from each chip convey the state of all checkers within the chip (Section II.E). The error monitors (Section II.F) then interpret these error signals and determine whether an attack has occurred.

Each chip implemented using TPAD includes four modules (Fig. 5): output encoding, input encoding, CED-based Trojan detection, and error encoding (Sections II.A-E).

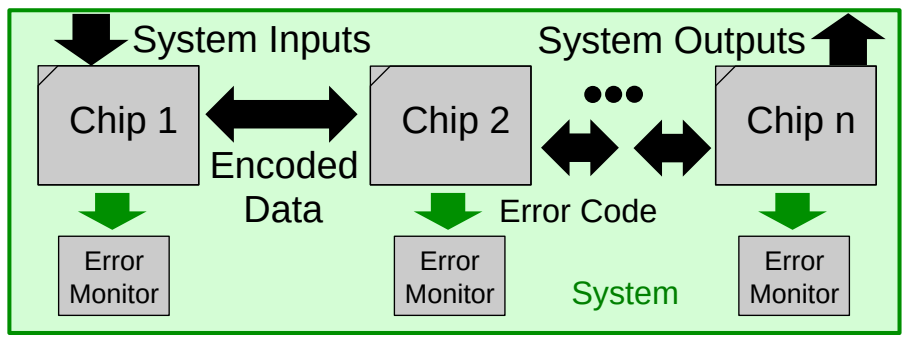

Fig. 4. Trusted system. Each chip is implemented with TPAD. Data communication between chips is encoded. Error monitors check encoded error signals and determine if an attack has occurred.

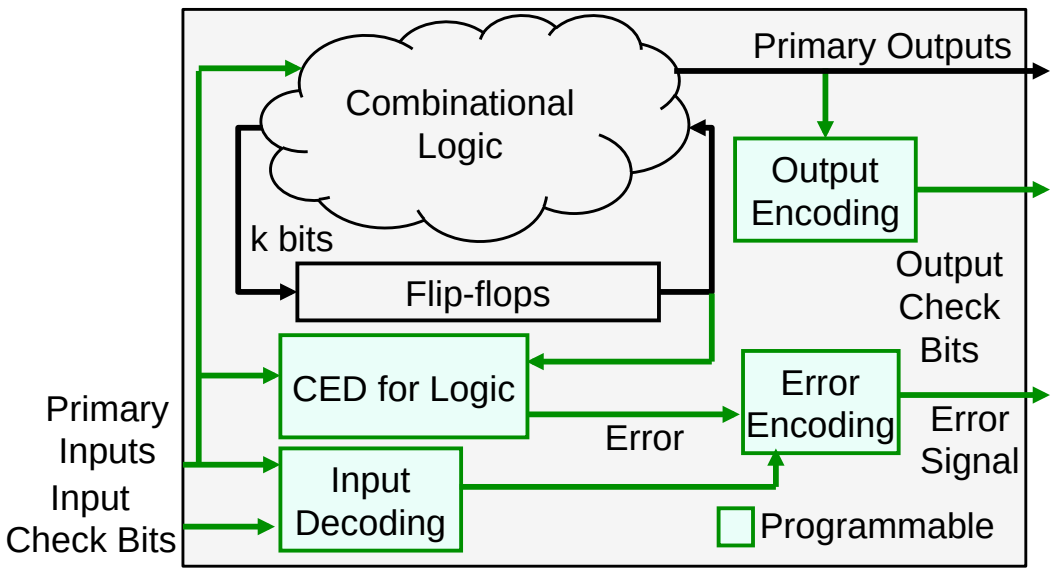

Fig. 5. TPAD architecture for each chip, including output encoding, input decoding, CED for logic, and error encoding.

## *A. Output encoding*

TPAD encodes primary outputs (Fig. 6a) using randomized parity encoding (Section IV.A). A separate encoding is used for each subsequent chip in the system that receives a different set of the primary outputs; the same encoding is used for chips receiving the same set of primary outputs (Fig. 6b). The primary outputs and their check bits can be transmitted to destination chips serially (same or different pins) or in parallel.

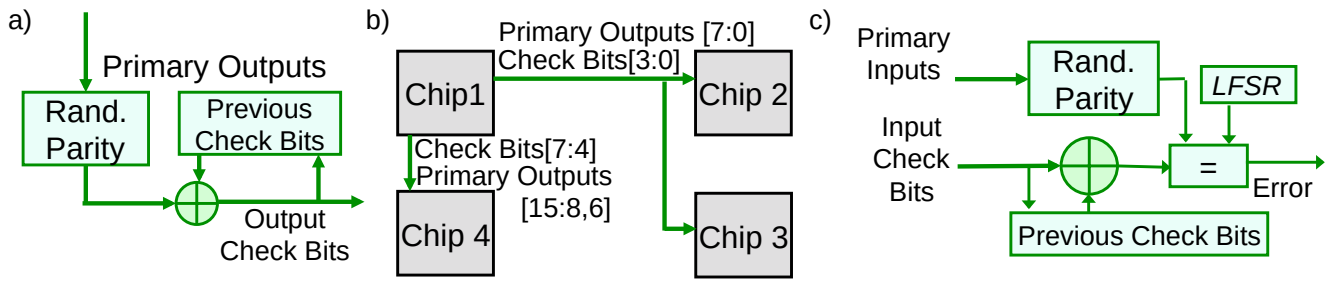

Fig. 6. a) Output encoding; b) Same encodings for the same subset of primary outputs; separate encodings for different subsets of outputs; c) Input decoding.

A randomized parity codeword (explained in Section IV.A) is calculated for the primary outputs during each clock cycle. The check bits of this codeword are then XOR'ed with the previous output check bits (stored in flip-flops), to form the output check bits (e.g., $110101_2 \oplus$ previous output check bits $010101_2$ produces output check bits $100000_2$). The output check bits are then stored and used in the next clock cycle. Thus, the output encoding at a particular time is a function of the history of the primary outputs in the preceding clock cycles (the starting check bits are initialized at chip startup to be uniformly random).

In order for a chip to check its primary inputs, it must use the same randomized parity encoding scheme as the sender's output encoding. To ensure this property, FIFOs or proper handshaking protocols may be required.

An attacker can attempt to derive the randomized parity scheme by adding hardware to the chip that stores randomized parity outputs and solves linear equations. However, even an attack that requires fewer additional gates can usually be detected using nondestructive post-manufacture inspections; thus, a complex attack such as this is expected to be detected (Appendix A).

## B. *Input decoding*

The primary inputs (Fig. 6c) of each chip are encoded according the method described in Section II.A. The decoding process checks for attacks at the outputs of the sender as well as attacks at its own inputs. Suppose that, primary inputs (e.g., $FA_{16}$) and input check bits (e.g., $1_{16}$) are received. The input check bits are then XOR-ed with the previous cycle's input check bits (e.g., $B_{16}$) to calculate the expected randomized parity bits (e.g. $1_{16} \oplus B_{16} = A_{16}$). The actual randomized parity bits are calculated from the primary inputs (e.g., $A_{16}$). Since both sets of randomized parity bits (expected and actual) are equal, no attack has occurred in this case. However, when a pin attack (Appendix A) occurs, the expected parity bits will not match the actual bits; thus, the pin attack will be detected.

## C. *Logic CED*

On-chip logic is protected using CED (Fig. 5) as introduced at the beginning of Section II. When the combinational logic is separated into independent blocks, different CED schemes may be used for each of the blocks as shown in Fig. 3b. A general CED technique for any digital system is discussed in Section IV, while application-specific techniques are discussed in Section V. These techniques overcome the limitations of classical CED in detecting Trojan attacks.

## D. *Memory CED*

Trojans inserted in an on-chip RAM (e.g. read/write logic, address decoder logic, and memory cells) can alter the data, the location in which the data is stored, or the operation of the RAM (read vs. write). To detect such attacks, the RAM is protected using a randomized parity code (Section IV.A). In TPAD, during a write operation, *both* the address (e.g., $BE_{16}$) and data bits (e.g., $124_{16}$) are used to calculate check bits (e.g., $6_{16}$) (Fig. 7a) to ensure that correct data is written to the correct location. These check bits are stored along with the data. During read operation, the address (e.g., $BE_{16}$) and data (e.g., $124_{16}$) are used to calculate the expected check bits (e.g., $6_{16}$). These are compared with the check bits read out from the memory (e.g., $6_{16}$) and an attack is detected if they do not match. For example, if the same data and check bits (as the above example) were retrieved during a read operation for a different address (e.g. $BF_{16}$), the expected check bits would not match the retrieved check bits (e.g. $2_{16} \neq 6_{16}$). To hide the randomized parity code construction from adversaries, both the encoder and checkers are protected with switchbox programmability (Section III).

For detecting attacks related to a write operation, the RAM must operate in write-though mode, a feature in many RAMs [25]. This means that during a write operation, the data that is being written is sensed and appears at the output of the RAM. Thus, immediately following a write operation, the input and output of the RAM block can be checked for attacks. Latches are used to ensure Data Out only changes during a read operation (Fig. 7a).

Table I shows the specific checking step that detects a Trojan for each operation (read, write, or idle). Fig. 7b shows the relative area of this Trojan detection scheme for various numbers of check bits as well as different RAM sizes. A word size of 16 bits (i.e. data in and data out were both 16-bits wide) was used for each case. GlobalFoundries 65nm SRAM compiler was used to obtain RAM areas. Checking circuits were synthesized and placed-and-routed using the GlobalFoundries 65nm CMOS library.

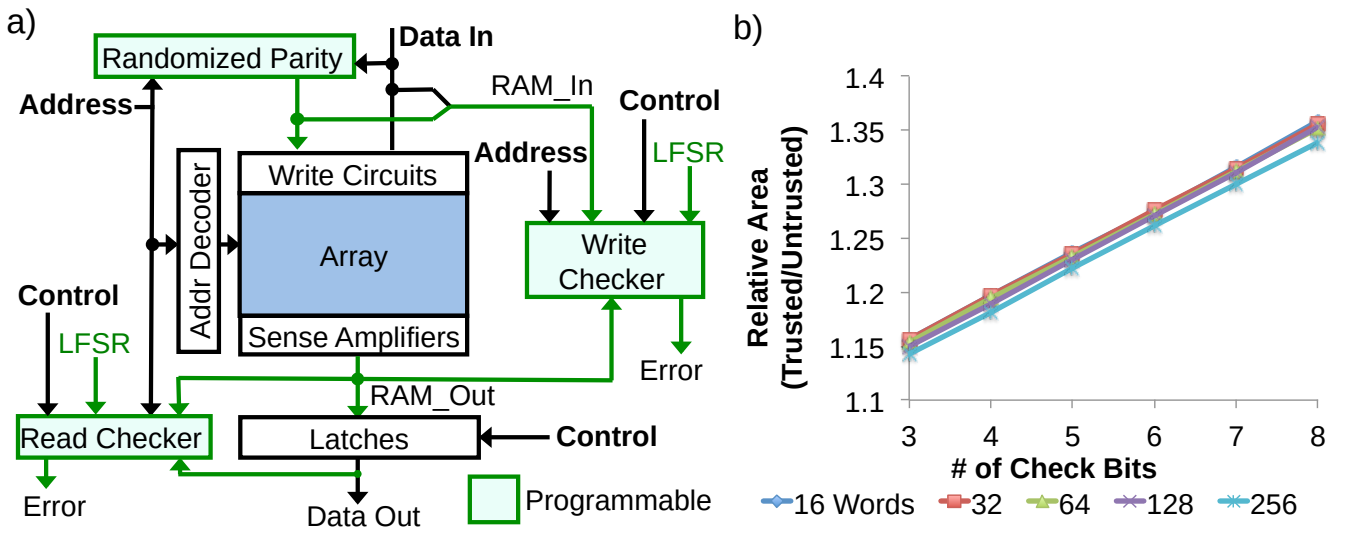

Fig. 7. Memory architecture. a) Block diagram of data encoding before writes and checking after reads/writes. Blocks with programmability indicated in green (Section III); b) Relative area cost of implementing trusted RAM compared to a non-trusted RAM vs. number of check bits.

TABLE I. POSSIBLE ATTACKS IN RAM AND TPAD DETECTION METHOD

| Operation | Effect of Trojan | Effect at Read/Write Checker |
|---|---|---|
| Read | Wrong address read | Check bits at RAM_Out incorrect |
| | Wrong data read | Check bits at RAM_Out incorrect |
| | Write instead of read | RAM_Out ≠ Data_Out |
| | Does not read | Check bits at RAM_Out incorrect |
| Write | Wrong write address | Check bits at RAM_Out incorrect |
| | Wrong data written | Check bits at RAM_Out incorrect |
| | Read instead of write | RAM_In ≠ RAM_Out |
| | Does not write | RAM_In ≠ RAM_Out. |
| Idle | Reads instead of idle | RAM_Out = Data_Out |
| | Writes instead of idle | RAM_In = RAM_Out |

## E. *Error signal encoding*

Error signals from various CED checkers are inherently vulnerable. For example, if a single bit is used to indicate that a Trojan attack has been detected, an adversary can simply attack that bit to indicate no attack.

Totally self-checking checkers [19] are also inadequate in detecting Trojan attacks, because an adversary can insert a Trojan into the checker that can make the checker output appear to be valid.

A uniform random sequence might be considered to prevent an adversary from guessing the meaning of the error signal since it has maximum entropy [26]. However, a deterministic error signal is needed so that it can be interpreted by the error monitor (Section II.F); thus, a Linear-Feedback -Shift-Register (LFSR) [27] is natural for this purpose. The polynomial and seed are made programmable to prevent an adversary from compromising the LFSR during design and fabrication. XOR gates are inserted in a shift register as shown in Fig. 8a to ensure that any primitive polynomial of degree $L$ can be realized. Programmability can be realized with RRAM switchboxes for low area cost (Section III).

Because of the large number of primitive polynomials for a sufficiently large degree (e.g., for $L$=64 it is on the order of $10^{17}$ [103]), the probability that an adversary correctly guesses which polynomial will be used during runtime is negligible. If the adversary is able to observe outputs of the LFSR during runtime, the evolution of the error signal may be deduced [28]. However, even an attack that requires fewer additional gates can usually be detected using nondestructive post-manufacture inspections; thus, a complex attack such as this is expected to be detected (Appendix A).

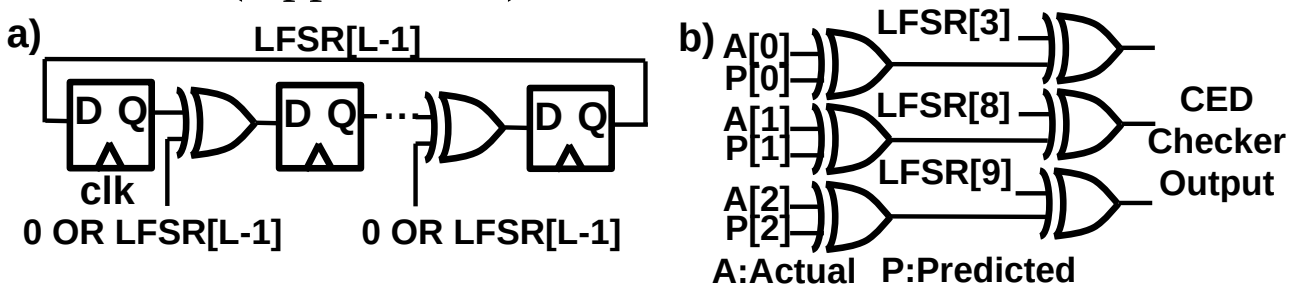

Fig. 8. a) LFSR of length L with programmable feedback polynomial. b) Sample CED checker design with $r=3$ bits.

A subset of $r$ bits in the LFSR of a given design (e.g. $35_{16}$), determined during design time, is used to encode the error signals for the CED checker. This subset can be chosen arbitrarily since the characteristic polynomial and seed of the LFSR are programmable. The checker must operate such that when no attack is detected, the error signal will be equal to the $r$ LFSR bits (e.g. $35_{16}$), shown in Fig. 8b. If the error signal takes any other value (e.g. $36_{16}$), an attack is detected. The subset of $r$ bits can be different for different chip designs.

All checkers within the same clock domain use the same LFSR. In any clock cycle when there is no attack detected, all of these checkers output the exact same signal. Thus, signals can be combined to reduce the number of output pins needed for the chip. To do so, each corresponding bit (e.g. bit 1 of checker 1, bit 1 of checker 2 etc.) is combined as shown in Fig. 9. When the LFSR bit is 0, the corresponding bit of the error signal is the OR of all of the checker outputs for that bit. When the LFSR bit is 1, the AND of the checker outputs is chosen. Checkers from different clock domains (integer and non-integer multiples) will have different LFSRs with proper FIFOs or handshaking to collect error signals.

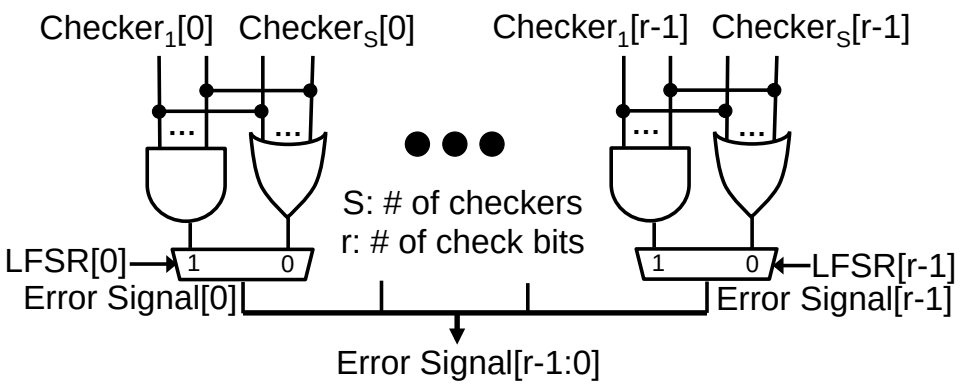

Fig. 9. Error signal encoding. For each of the r bits, if LFSR value is 0, the OR of corresponding checker output bits is chosen. If it is 1, the AND is chosen.

### *F. Error monitors*

Error signals must be interpreted by a trusted party to ultimately decide whether an attack has occurred. Separate LFSRs are used within each error monitor to generate expected error signals. These LFSRs are configured the same (characteristic polynomial, seed, and clock) as the LFSRs that generate the corresponding error signals. The error signal received should match the expected error signal; otherwise, an attack has occurred. To realize this error monitor, an older technology node from a trusted foundry may be used.

## III. Switchbox Programmability

As discussed in Section II, the output encoding, input decoding, logic CED, and error encoding blocks in Fig. 5 must be protected against modifications that stop Trojans from being detected. TPAD builds on ideas from privacy techniques in social networks [29] in order to prevent circuit reverse-engineering and sophisticated attacks. We represent logic gates in a netlist as vertices in a graph, and wires as directed edges between vertices. The netlists corresponding to the output encoding, input decoding, logic CED, and error encoding blocks in Fig. 5 are modified with a technique that prevents adversaries from gaining detailed knowledge of their functionalities. The technique has two important features:

1. Each modification to a netlist will require adding an extra logic element (Definition 1). The area and power cost of each such element is made small via the use of an emerging memory technology (discussed in this section).
2. The total number of modifications (overall cost) to a given netlist is flexible, and can be tuned based on the level of security desired. We show (Section III.A-V) that one can ensure exceptionally high security against a very wide variety of attacks while maintaining low total cost.

While the adversary can still bypass the checkers by storing a previous state of the system with additional hardware, these attacks can be easily detected using non-destructive post-manufacturing inspections (Appendix A). TPAD uses *random switching* [30] to hide functionality of checkers.

**Definition 1.** Let $e_1 = (x, z)$, and $e_2 = (y, w)$ be two disjoint wires (i.e., $x \neq y \neq z \neq w$). We say $e_1$ and $e_2$ are *switchable* if they can be reconfigured as $\tilde{e_1} = (x, w)$ and $\tilde{e_2} = (y, z)$.

This can also be generalized to larger sets of wires. Let

$$S = \{(a_1, b_1), (a_2, b_2), \dots, (a_k, b_k)\}$$

be any arbitrary set of $k$ disjoint wires. We say the wires in $S$ are *switchable* if for any permutation $\pi: [k] \rightarrow [k]$, $S$ can be reconfigured as $\tilde{S} = \{(a_1, b_{\pi(1)}), (a_2, b_{\pi(2)}), \dots, (a_k, b_{\pi(k)})\}$.

We refer to any circuitry that makes wires switchable as a *switchbox (SB)*. The total number of possible states of all the SBs in a circuit is referred to as the *number of configurations* of the circuit. An example of modifying a circuit using a SB is illustrated in Fig. 10. The circuit in Fig. 10a acts as a one-bit full-adder when the SB is configured as shown. However, if the SB is configured as shown in Fig. 10b, the circuit computes a different function. Thus, when certain wires in a design are made switchable, attackers must know the intended SB configuration to deduce the intended functionality.

However, simply inserting a SB into a design does not ensure that an incorrect configuration will lead to incorrect functionality. For example, a *degenerate* case of SB insertion is shown in Fig. 11. Regardless of the configuration of the SB, the circuit will behave as a full-adder. For this reason, our technique for inserting SBs involves formal equivalence checking to verify that each SB that is inserted can lead to different functionality (Step IV of Algorithm 1).

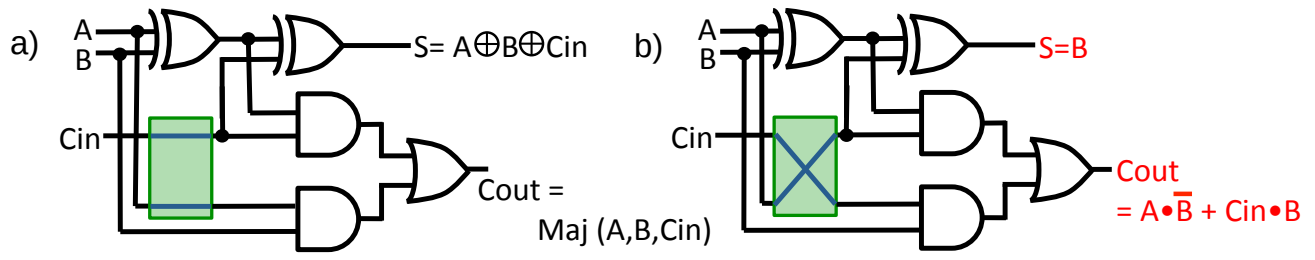

Fig. 10. Example of modifying a full-adder with a two-input switchbox. There are two possible configurations: a) parallel mode or b) crossed mode.

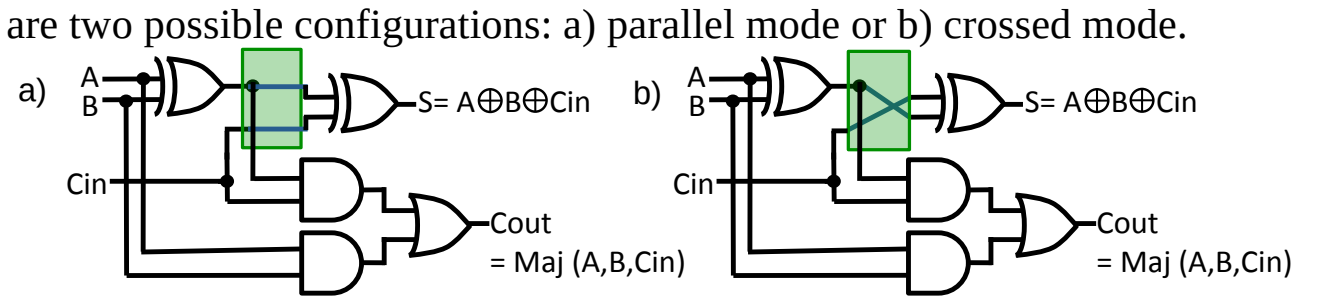

Fig. 11. Degenerate SB insertion in a full-adder with a two-input switchbox. Both configurations of the SB yield the same functionality.

However, degenerate cases can still occur when a larger number of SBs are inserted into a design. For instance, if $x$ two-input SBs are inserted into a design, there are $2^x$ total configurations of the SBs. If $x$ is small, it is feasible to run equivalence checking to verify that *only one* of the $2^x$ configurations yields the intended functionality of the design. However the number of equivalence checks needed to verify this property grows exponentially in $x$, making the problem of guaranteeing no degenerate cases intractable for large $x$.

To simulate percentage of degenerate cases for a large number of SBs, 64 SBs were inserted in an OpenRISC [39] instruction fetch module. $10^6$ incorrect SB configurations were applied and equivalence checking was performed to compare the resulting functionality with the intended design. No degenerate cases were found. We note however, that only a small fraction (0.05%) of the $2^{64}$ different configurations were explored in these simulations. Therefore a larger experiment would need to be performed in order to verify that degenerate cases are highly unlikely for a given design. The percentage of degenerate cases also depends on symmetry properties of the function being covered, and the SB insertion technique would need to account for such properties in order to provide guarantees.

If an SB insertion technique ensures the realization of multiple checking function instances using the same set of SBs, the presence of degenerate cases can be circumvented. For example, if one could insert SBs into a logic CED checker in such a way that *many* CED techniques, e.g., exponentially many randomized parity codes of a given length (Section IV.A), can be realized, then it would be very difficult for the adversary to infer which particular CED technique would be used during runtime. One way of ensuring multiple CED techniques is to insert SBs in the OCP and the checker *before* logic synthesis such that the randomized parity functions may be reconfigured. However, this can result in much of the original function being retained in the OCP, making the design vulnerable to certain sophisticated attacks (discussed in Section III.A). These vulnerabilities can allow an adversary to insert undetectable Trojans in those parts of the OCP without any knowledge of the CED technique being used. Thus, the SB insertion method described in this section still must be used *after* synthesis to ensure an adversary cannot reverse-engineer the OCP and the checker.

In hardware, a low area-cost technique for switching sets of wires is enabled through the use of Resistive RAM (RRAM) technology. Advances in RRAM technology have shown the feasibility of monolithically integrating RRAM with CMOS [32]. Its cell size has been shown to be smaller than other NVRAM technologies at $4F^2$ (F is the feature size of the technology node) [33]. Additionally, it has been used recently to secure memory designs against unauthorized read accesses [31]. RRAM provides a non-volatile configuration memory for SBs so chips do not have to be reconfigured when the power is cycled, making chip boot-up more secure since there is no need for potentially insecure programming sequences to be sent to the chip. Moreover, when the design rarely requires reprogramming, long write times of the RRAM are negligible.

Fig. 12 illustrates the design of an RRAM SB as described and fabricated in [32]. Each cell consists of two RRAM elements arranged as a resistive divider, a programming transistor, and a pass transistor. The two RRAM elements in each cell are always programmed to be in different states, thus pulling the pass transistor gate either high or low, allowing or denying a given input to propagate to an output. The address bus (addr), program data in (prg_in), and write enable (we), are only used when configuring the SB, making the programming interface much like an SRAM interface. The SBs can be implemented with as little as 0.04% area overhead compared to the rest of a chip (Section V.A) since the RRAM cells can be placed in the metal interconnects above the active layer of transistors on a chip as described in [102]. Alternatively, standard technology such as flip-flops or lookup tables (LUTs) can be used to configure SBs, though non-volatility would be sacrificed and the area cost would be significantly higher.

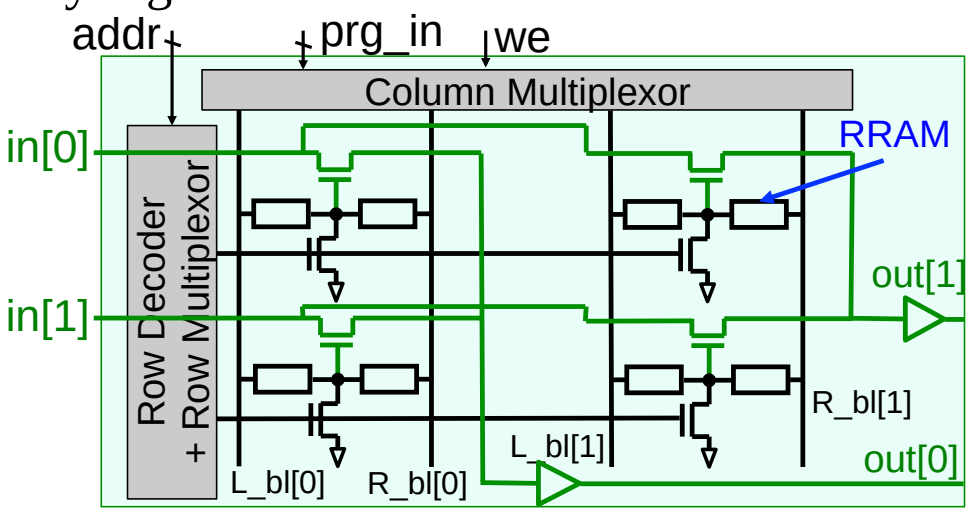

Fig. 12. Schematic of RRAM-based switchbox [32]. Address bus (addr), program data in (prg_in), and write enable (we) are used to program the switchbox to a particular configuration.

Algorithm 1 gives a method for randomly inserting switchboxes into any netlist, given the target number of configurations as an input. Fig. 13 illustrates the outcomes of the steps of Algorithm 1, when performed on the circuit shown in Fig. 13a, with 4 as the number of configurations per output bit. Steps I-II are illustrated in Fig. 13a and Steps III-V are illustrated in Fig. 13b. The effectiveness of this algorithm in preventing sophisticated attacks is evaluated in Section III.A.

While it is possible to insert SBs into the original functions instead of the checking circuitry, this leaves the entire end-to-end CED scheme visible and it can be modified to never detect errors in the original functions. Thus, a trusted end-to-end CED scheme is imperative for detecting Trojans. When split-manufacturing is used, SB programmability is not needed to prevent attacks that require detailed knowledge of the design. The area savings with split-manufacturing are discussed in Section IV.B - V.A.

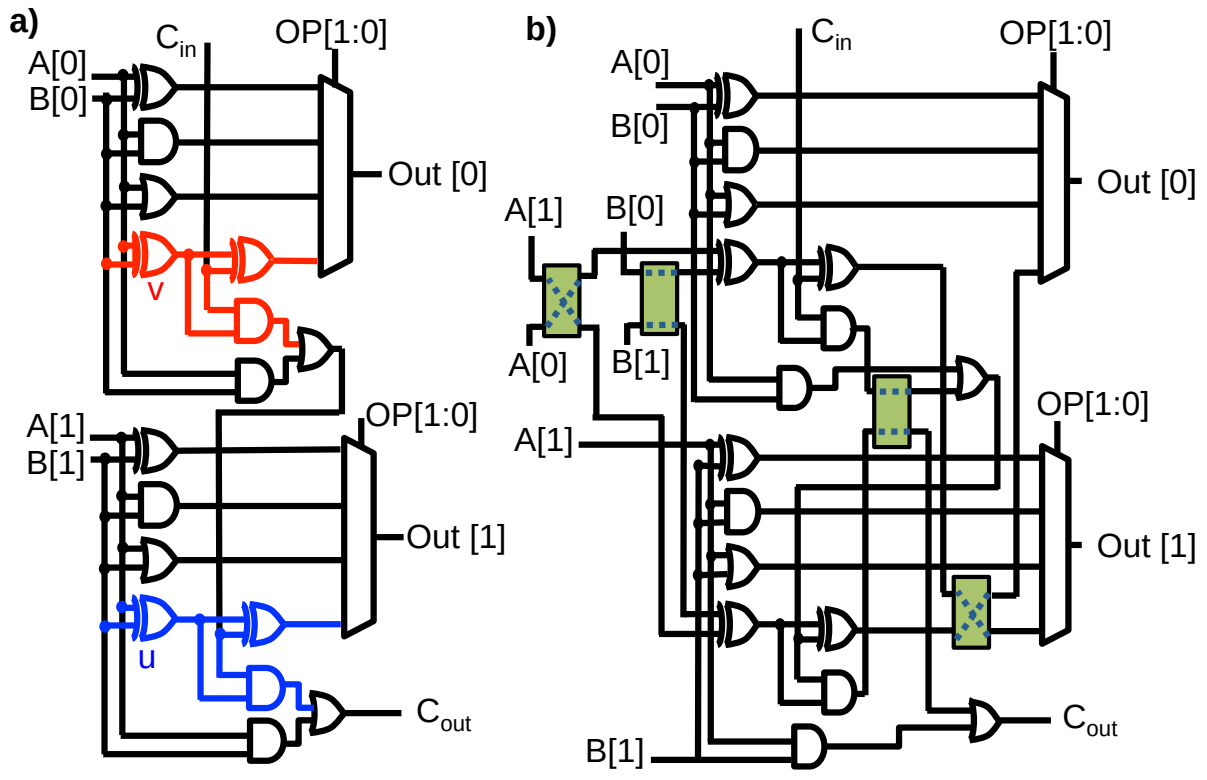

Fig. 13. Steps of Algorithm 1 applied to a 2-bit ALU targeting 4 configurations per output bit; a) Step I: choose v and its neighborhood (highlighted in red); Step II: choose u and its neighborhood (highlighted in blue); b) Step III: Place SBs (shown in green); Step IV: Test that SBs are not degenerate (satisfied) and choose configurations (shown in dotted lines); Step V: Test for at least 2 SBs within the logic cone of each output bit (satisfied).

**Algorithm 1 [Random Switchbox (SB) Insertion]**
**Inputs:** Netlist $G = \{V, E\}$, Number of configurations per output bit: $2^t$.
**Outputs:** New netlist $\tilde{G} = \{\tilde{V}, \tilde{E}\}$ with $\geq 2^t$ configurations per output bit.
Initialize: $\tilde{G} = G$.
Step I: Choose $v \in V$ uniformly at random and group all the vertices within radius 1 of $v, N_1(v)$.
Step II: Search $\tilde{G}$ for $u \in V$ s.t. in-degree and number of outputs of $N_1(u)$ match $N_1(v)$, and $N_1(u) \cap N_1(v) = \emptyset$. If not found, retry Step I.
Step III: Let $N_1(v)$, $N_1(u)$ each have $a$ incoming wires $\{(x_1, z_1), \ldots, (x_a, z_a)\}$ and $\{(y_1, w_1), \ldots, (y_a, w_a)\}$, and $b$ outputs $\{s_1, \ldots, s_b\}$ and $\{t_1, \ldots, t_b\}$ respectively.
**For** $1 \leq i \leq a$:
Insert two-input SB in $\tilde{G}$ with inputs $(x_i, y_i)$, outputs $(z_i, w_i)$.
**For** $1 \leq j \leq b$:
Insert two-input SB in $\tilde{G}$ with inputs $(s_j, t_j)$, first output for all outgoing edges of $s_j$, second output for all outgoing edges of $t_j$.
Step IV: **For** each SB inserted in Step III:
Consider $G$ with only this SB added.
Compare functionality with this SB crossed vs. this SB parallel.
If equivalent: Remove this SB.
Else: With probability ½, adjust input and output wires of SB so that crossed SB will yield same functionality as $G$.
Step V: Let $G_i$ denote the subgraph of $\tilde{G}$ corresponding to the logic cone which ends at the $i$-th output bit of $G$. If $\forall i$, there are at least $t$ SBs in $G_i$, output $\tilde{G}$. Otherwise repeat steps I-IV.

## *A. Resilience to reverse-engineering and sophisticated attacks*

In this subsection, we provide results showing the effectiveness of SB programmability in preventing attacks that require some detailed knowledge of the design. First, we state an assumption that is used to set the number of configurations input to Algorithm 1 when it is applied in Sections IV.B-V.

**Assumption 1.** If an adversary learns the definition of *any* characteristic function computed by *any* attack detection circuit (e.g., Logic CED, output encoding, input decoding), a Trojan can be inserted into the design that cannot be detected.

For example, assume an OCP with two check bits is designed for a function with four outputs. Assume the first check bit is used to predict the parity of output bits 1 and 2, while the second check bit is used to predict the parity of output bits 3 and 4. If an adversary figures out the first characteristic function and flips the first two bits at the output of the main function, this attack would not be detected.

Assumption 1 is used in Step V of Algorithm 1. The Trojan detection circuitry is decomposed into separate logic cones for each output bit. Then, programmability is selectively inserted into each logic cone in order to hide the function that is used to compute each output bit.

The following list states reverse-engineering attacks (e.g., attempts to gain some detailed knowledge about the intended functionality of a design) and how they are prevented by SB programmability. Each attack was simulated on an OpenRISC CPU protected with randomized parity codes (Section IV). The probability that the attack succeeded is shown in Table I. An attacker can try all possible SB configurations or examine the physical aspects (i.e., gates and/or wires) of the circuit or sub-circuits. A more general attack model (Configuration Prediction Attack) assumes the attacker can determine the correct configuration for each SB with a given probability. Even if this probability is high, the overall probability of reverse engineering the circuit can still be made negligible by inserting a large number of SBs.

1. Brute Force Attacks: The attacker isolates a single output bit in the netlist, and searches through all possible SB configurations along the logic cone leading to it. For example, the adversary may configure the SBs in all possible ways, and create a list of all possible functionalities. Then the adversary may guess the intended configuration. We prevent these attacks by inserting a *large number* of SBs within *each* logic cone leading to an output bit in *all* of the Trojan detection circuits. For example, in Section IV.B, we insert between 64 and 80 SBs along every logic cone in the Trojan detection blocks of a CPU. The number of SB configurations of each cone (ranging from $2^{64}$ to $2^{80}$) is too many to exhaustively test in order to reverse-engineer the functionality.
2. Configuration Prediction (CP) Attacks: A $\langle\theta\rangle$ CP attack assumes that the adversary can correctly guess the intended configuration of any individual SB with probability $1-\theta$. Since the intended configuration of each SB is chosen independently of the others, the minimum-bias strategy to guess the configurations of all the SBs in a circuit is to guess each SB configuration independently [105]. Hence, if there are $x$ SBs within the logic cone leading to a characteristic bit, a $\langle\theta\rangle$ CP attack assumes an adversary can guess the configuration of that logic cone correctly with probability $(1-\theta)^x$. We prevent such attacks by inserting a *large number* of SBs within *each* logic cone leading to an output bit in *all* of the Trojan detection circuits, causing these attacks to succeed with negligible probability (i.e., $x$ is made large so that $(1-\theta)^x$ will be small; Table II).
3. Wire-length Tracing: If a physical design tool was given the intended configurations of the SBs, wire-lengths for the intended SB configurations could be optimized to minimize delay, while wire-lengths for incorrect configurations could be ignored. Then, an adversary could compare the wire-lengths of both configurations for each SB and find the one resulting in shorter wires. We prevent this attack by *not providing* the tool with the intended configurations (Section VI). Thus, no single SB configuration is optimized at the expense of others during physical design (Fig. 14a and Table II). To demonstrate this, SBs in the parallel configuration were inserted in various OpenRISC OCP designs (Section IV) with the intended configuration undisclosed to the physical design tool. The difference in the maximum wire-lengths for parallel ($L_{parallel}$) and crossed ($L_{cross}$) configurations was determined for each SB. If $L_{parallel} < L_{cross}$, the adversary would guess parallel; otherwise, the adversary would guess crossed. The median value (out of 2,500 trials) of $L_{parallel} - L_{cross}$ was 0, indicating that an adversary would guess correctly with probability 0.5. Thus, this attack provided no advantage over random guessing
4. Drive Strength (DS) and Fanout (FO) Matching: If SBs were inserted *post-physical design*, an adversary would be able to match (strong or weak) driving cells at the inputs of SBs to (large or small) FOs at the outputs. For example, let SB inputs be driven by gates with DS 1x and 4x, and SB outputs have fanouts 1 pF and 4 pF. An adversary can guess that the 4x cell was meant to drive the 4 pF load to minimize delay, thus revealing the SB configuration. We prevent this attack by inserting SBs *before physical design* and using SBs with equal capacitances at the input ports. Since physical design will size both driving cells to drive the SB, the DS of both inputs will be similar, regardless of FO. This attack was implemented 4,000 times on various OpenRISC OCP designs. Random guesses for the SB configuration were used to break ties. As shown in Fig. 14b, this attack provided only a 0.01% increase in success rate when compared to random guessing.

5. Subcircuit Matching: An adversary can search for isomorphic subcircuits shared by a function and its OCP. Since the OCP is a composition of characteristic functions with the main function, isomorphic subcircuits may be used to compute the same logic values in both blocks. An attack that flips the outputs of both subcircuits would not be detected by the CED tests, since check bits would still match. SB insertion breaks up logic cones in the OCP, leading to far fewer isomorphic subcircuits. Equal modification to the remaining isomorphic subcircuits is detected with high probability, indicating that even though they are constructed with the same logic gates, they are not used to compute the same logic values (Table II). This attack was implemented on the OpenRISC CPU (Section IV) 40,000 times and was unsuccessful every time.

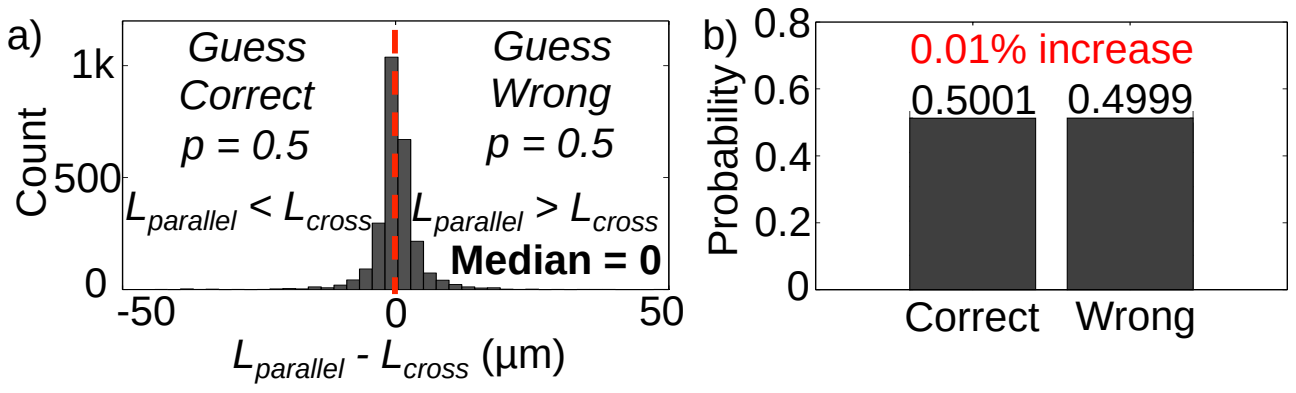

Fig. 14. a) Wire-length tracing results. The attack succeds 50% of the time. b) Drive Strength and Fanout Matching attack results. Using this attack resulted in only a 0.01% advantage in success rate over random guesing.

TABLE II. REVERSE-ENGINEERING SUCCESS PROBABILITIES

| Type of Attack | Configurations/Output bit | | |
|---|---|---|---|
| | $2^{64}$ | $2^{72}$ | $2^{80}$ |
| Wire-length Tracing* | $5.4\times10^{-20}$ | $2.1\times10^{-22}$ | $8.3\times10^{-25}$ |
| DS & FO Matching* | $5.5\times10^{-20}$ | $2.1\times10^{-22}$ | $8.4\times10^{-25}$ |
| Subcircuit Matching* | 0.0 | 0.0 | 0.0 |
| CP: $\theta = 0.1$ | $1.18\times10^{-3}$ | $5.08\times10^{-4}$ | $2.18\times10^{-4}$ |
| CP: $\theta = 0.05$ | 0.0375 | 0.0249 | 0.0165 |

* Simulated using OpenRISC w/ Randomized Parity codes (Section IV).

## IV. RANDOMIZED CODING FOR LOGIC CED AND SECURE I/OS

### A. Randomized parity codes

We assume the reader has familiarity with the basics of coding theory [24] and its use in fault-tolerant computing [19]. We provide only the minimum definitions here. The notation used is defined in Table III. All the presented codes are binary, but they can be extended to larger finite fields.

TABLE III. NOTATION

| | |
|---|---|
| $\{0,1\}^n$ | Binary strings of length $n$ |
| $x_i, i \in \mathbb{Z}$ | $i$th element of vector $x$. |
| $x^T$ | Transpose of vector $x$. |
| $\bigoplus$ | Modulo-2 sum (XOR). |
| $\mathbb{F}_2$ | Binary Field: $\{0,1\}$ with $\bigoplus$, mod-2 multiplication |
| $M^{n\times k}(\mathbb{F})$ | Set of $n\times k$ matrices over field $\mathbb{F}$. |

**Definition 2.** A $k$-dimensional *parity function* $g:\{0,1\}^k \to \{0,1\}$, is a parity of a subset of bits of a k-dimensional binary vector. Explicitly, $\forall x \in \{0,1\}^k$:

$$g(x) = \bigoplus_{i=1}^{k} a_i x_i \quad a_i \in \{0,1\}.$$

**Definition 3.** An $(n,k)$ *parity (linear) encoding* $g:\{0,1\}^k \to \{0,1\}^n$ is a mapping where each coordinate function is a $k$-dimensional parity function. We call $n$ the *blocklength* of the code, and $k$ the number of *information bits*.

**Definition 4.** A *parity-check matrix* $H \in M^{r\times n}(\mathbb{F}_2)$ for an $(n,k)$ linear encoding $g$, is a binary matrix for which $Hx^T = 0$ iff. $\exists y \in \{0,1\}^k$ s.t. $g(y) = x$ . $C \subseteq \{0,1\}^n$ with $C = \{x \in \{0,1\}^n \mid Hx^T = 0\}$ is called the *code*. The elements of $C$ are called the *codewords*. A linear code is *systematic* if its parity-check matrix has form $[A \mid I_r]$ where $I_r$ is the $r\times r$ identity matrix and $r$=$n$-$k$ is the number of *check bits*.

Fig. 15 illustrates how linear systematic codes are used in concurrent error detection [34]. The information bits of a codeword are computed by a logic function. The OCP predicts the corresponding check bits, and the checker computes the actual check bits and compares them with the prediction. We now introduce a randomized linear code construction.

**Construction 1** (Randomized Parity Codes). Let $R^{n\times k}$ denote the ensemble of all $(n,k)$ linear systematic codes with all rows and all columns in the parity-check matrix nonzero. An $(n,k)$ *randomized parity code* is a code that has been sampled uniformly at random from $R^{n\times k}$.

Randomization of the code choice, in conjunction with SB programmability (Section III), prevents an adversary from knowing the parity-check matrix. The nonzero constraint on each row of the parity-check matrix is to avoid the zero function. This function is useless for error detection, since it corresponds to a permanently grounded wire. The nonzero constraint on the columns is to ensure that each input bit is included in at least one parity function. Algorithm 2 constructs a randomized parity code.

**Algorithm 2 [Randomized Parity Construction].**
**Inputs:** Number of information bits $k$. Number of check bits $r$.
**Output:** Parity-check matrix for an $(r+k, k)$ randomized parity code.
Step I: Choose a systematic parity-check matrix $H \in M^{r\times(r+k)}(\mathbb{F}_2)$ uniformly at random.
Step II: If $H$ satisfies constraints of Construction 1: output $H$.
Else: repeat Steps I-II.

For any given error pattern, a code chosen uniformly at random from the ensemble of *all* (n,k) binary linear systematic codes will detect the errors with probability $1 - 0.5^r$. However, since Construction 1 introduces dependencies between columns in the parity-check matrix, the standard proof of the error detection probability will not hold [37]. Fig. 15 therefore shows the detection probability for randomized parity codes, for $k$=100, simulated using Monte Carlo (30,000 trials). The detection probability is 1 for single-bit errors and approaches $1 - 0.5^r$ as the number of errors increases [38]. If a nonlinear code is used instead of a parity code, the detection probability for a large number of errors can be higher at the cost of additional area [35].

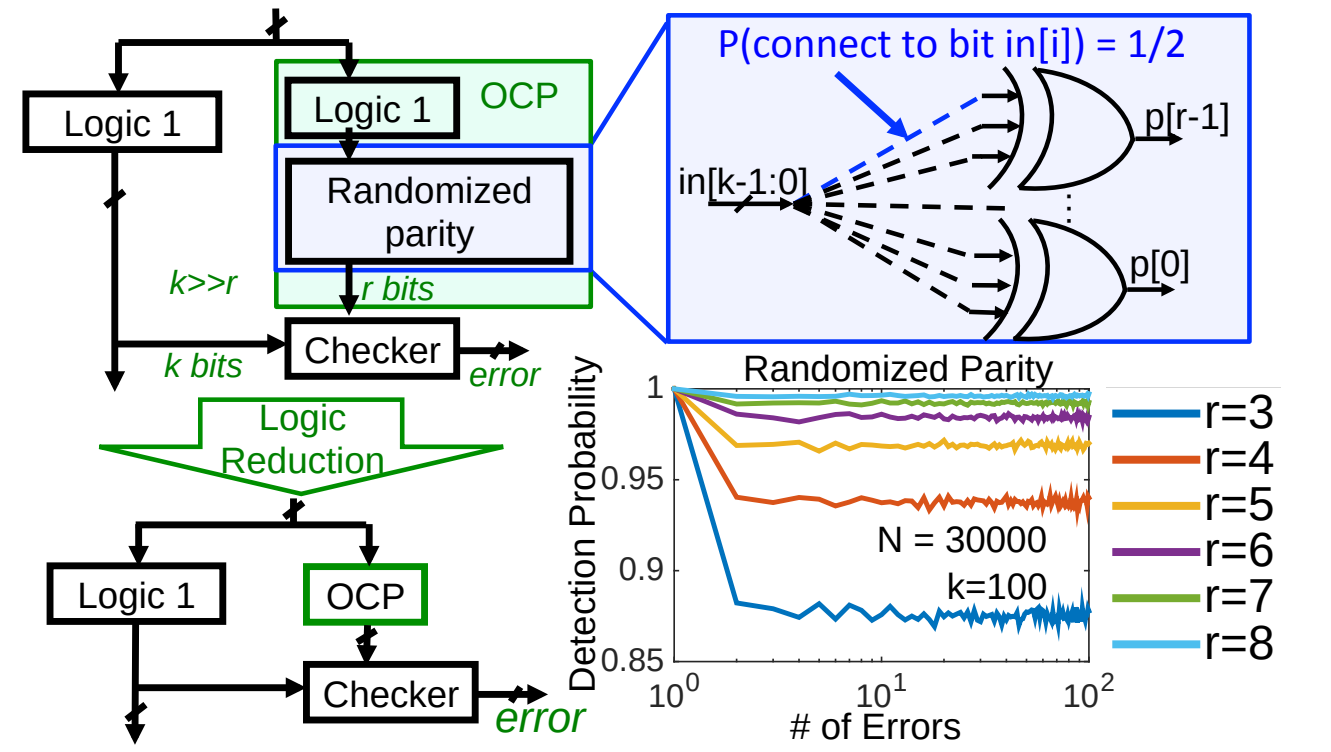

Fig. 15. General method to create an OCP. Detection probability for various number of check bits for *k*=100 are shown for randomized parity codes.

## *B. OpenRISC CPU demonstration*

To evaluate TPAD, we constructed an OpenRISC 1200 CPU core [39] using randomized parity code checking (Section IV.A) for logic CED, input decoding, and output encoding. Designs were synthesized and place-and-routed with the Nangate 45nm CMOS standard cell library [40]. The number of check bits for each pipeline stage ranged from 3 to 8 bits, and SB programmability (Section III) was included for all checking circuitry and I/Os. The same number of check bits was used for all pipeline stages within the processor. The relative size and power of the overall design for 3 check bits (which corresponds to a detection rate of 87.5%) is given in Fig. 16. This shows an area overhead of 114% and power overhead of 34%. The number of SB configurations used for each bit of both the OCP and characteristic function in Algorithm 1 was $2^{64}$. This corresponds to a delay increase of 20% (the critical path is now in the OCP) if the same number of pipeline stages is used. However, this delay overhead can be reduced to 0 if an extra pipeline stage is added in one of the OCPs and its corresponding checker. The area and power overheads of this change were found to be negligible. Additional results for $2^{72}$ and $2^{80}$ SB configurations for each OCP bit are given in Table IV. To simulate the detection probability, a uniformly random number of bit flips were introduced to the system at uniformly random times.

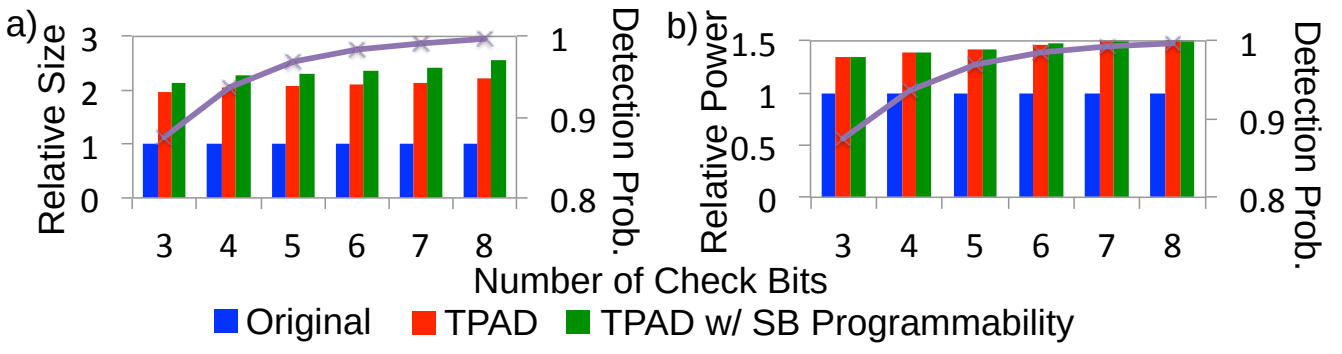

Fig. 16. Area and power overheads and detection probability for secure OpenRISC 1200 processor design using randomized parity codes with $2^{64}$ SB configurations per check bit for all OCPs and checkers.

Most of the area overhead of randomized parity coding is due to the inefficient logic reduction of the OCP. This results in a significant portion of the OCP resembling the original function. While this seems to imply that the chip is vulnerable to sophisticated attacks (i.e., an adversary can make the same change to the original function and OCP), SB programmability prevents such attacks by breaking up logic cones in the OCP (Section III). The logic reduced and pre-logic reduced forms of the OCP for all OpenRISC modules were compared to see how many wires corresponding to the inputs to the parity functions (i.e. *in[k-1:0]* in Fig. 15) were retained. More than 99% of these nets were found to be reduced, rendering finding full logic cones from the original circuit difficult. Although SB programmability creates additional overhead, the marginal area (17-33%) and power (0.4%) costs are low compared those of the OCP functions.

TABLE IV. AREA, POWER, & DELAY OVERHEAD TRADEOFFS FOR OPENRISC

| *r* | Overheads for # of SB configurations / OCP bit: $2^{64}$ Area | Power | Delay | $2^{72}$ Area | Power | Delay | $2^{80}$ Area | Power | Delay | Logic Attack Detection Probability |
|---|---|---|---|---|---|---|---|---|---|---|
| 3 | 114 | 34 | 0 | 122 | 34 | 0 | 121 | 34 | 0 | 0.875 |
| 4 | 127 | 39 | 0 | 130 | 39 | 0 | 131 | 39 | 0 | 0.937 |
| 5 | 131 | 43 | 0 | 135 | 43 | 0 | 139 | 43 | 0 | 0.968 |
| 6 | 137 | 47 | 0 | 142 | 47 | 0 | 145 | 47 | 0 | 0.984 |
| 7 | 142 | 53 | 0 | 146 | 53 | 0 | 150 | 53 | 0 | 0.992 |
| 8 | 157 | 60 | 0 | 158 | 60 | 0 | 165 | 61 | 0 | 0.996 |

## *C. Secure FPGA demonstration*

To demonstrate TPAD in FPGAs, an FPGA using randomized parity coding to secure the I/O signals was fabricated in 65nm CMOS technology. Since the FPGA is programmable (discussed in Section II), logic CED can be mapped to the CLBs to protect the FPGA-mapped logic.

However, as discussed in Section II, FPGAs can still be vulnerable to attacks. This FPGA does not contain any accelerator blocks, but the I/O ports of the FPGA still must be protected against pin attacks (Appendix A). Additionally, the memory row decoder needs to be protected. The inputs to the FPGA are encoded with a randomized parity code. This encoding is checked on every clock cycle for errors (Fig. 17).

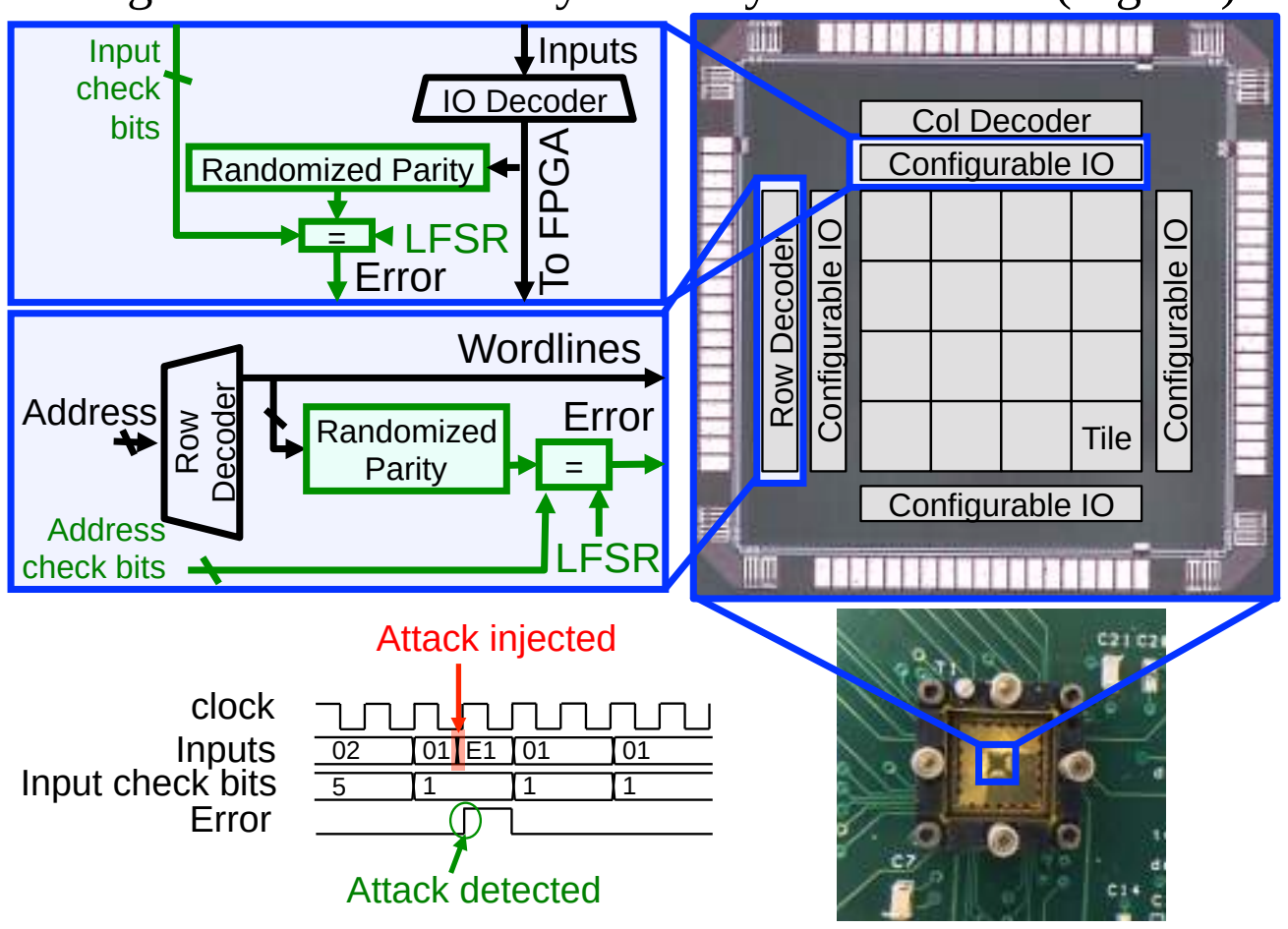

Fig. 17. Pin attack protection for FPGA. Chip in testing board, die photo, and block diagram shown. An example of an emulated pin attack is also shown.

The design uses 3 check bits for each of the 4 I/O blocks and an additional 3 bits for the memory row decoder, yielding an area overhead of 15% and a power overhead of 0.07% with no critical path delay change. Pin attacks were emulated at the inputs to the chip by selecting random bits of the input stream to be flipped. The checker was monitored to see if the attack was detected. The detection rate for pin attacks was 87.5%, matching the theoretical prediction. Area, power, and delay overheads and detection probability for both circuits (OpenRISC CPU and FPGA) are summarized in Table V.

TABLE V. RESULTS FOR TPAD WITH RANDOMIZED PARITY

| Area/Power/Delay Overhead (%) | | Attack/Error Detection Probability | | |
|---|---|---|---|---|
| Circuit: | TPAD | Pin | Logic | Random Error |
| OpenRISC | 157/60/0 | 0.996 | 0.996 | 0.996 |
| FPGA | 15/0.07/0 | 0.875 | 1.00 | 0.875 |

## V. OVERHEAD REDUCTION VIA SPECIALIZED OCP FUNCTIONS

While randomized parity for logic CED offers high Trojan detection probability for arbitrary designs, it does not necessarily offer the best area overheads (similar to application-specific trade-offs for fault-tolerant computing [19]). Application-specific methods can possibly decrease the area overhead. However, since sophisticated Trojans are able to escape many checks (as discussed in Section II), application-specific CED techniques for fault-tolerant computing cannot be used unmodified for TPAD. For example, sum of squares (a common CED technique for Fast Fourier Transform (FFT) circuits [36]) cannot detect any attack that permutes (in any arbitrary way) the FFT outputs.

### *A. LZ data compression engine*

For certain invertible functions, the inverse is simpler to implement than the forward function. In such cases, TPAD can use the inverse to detect Trojans efficiently. We demonstrate this by implementing a Lempel-Ziv (LZ) data compressor chip. The chip uses the sliding-window version of the LZ data compression algorithm [45] (i.e. LZ77). The CED scheme used to protect the encoder is described in [23]. A LZ77 codeword is represented as a fixed length binary tuple $(C_p, C_l, C_n)$ where $C_p$ is the pointer to the location in the dictionary, $C_l$ is the match length and $C_n$ is the next character.

To detect attacks, the LZ codeword is decompressed by a dictionary lookup and compared with a copy of the input stored in a buffer. If there is any mismatch, an error has occurred. Because LZ77 is a lossless compression algorithm, changing a codeword to any other word will result in errors being detected. This inverse function method alone, however, is not enough to prevent an adversary from inserting Trojans. For instance, an adversary can make the same change to the output of the inverse function and the output of the input buffer to make the compared words match. Thus, SB programmability realized with RRAM (Section III), is needed for the CED. Moreover, output encoding and input decoding for the I/Os of the chip are needed to protect against pin attacks.

Since the dictionary keeps a record of past inputs, this CED method will detect not only any logic attack, but also decoupling attacks (Appendix A) in the compressor since the pointers to each dictionary entry change after every cycle. Thus, a previous LZ77 codeword will no longer point to the same entry in the dictionary.

A block diagram and die photo of our split-manufactured test chip, implemented in 65nm technology, with secure I/O encoding (Section II.A-B) is shown in Fig. 18. To emulate logic attacks, a number of random flip-flops inside the chip were flipped at random times using a scan chain. To emulate electrical attacks, flip-flop values were reverted to their previous value (setup time violation) or changed to their next value (hold time violation). To emulate reliability attacks, internal wires were forced high or low at random times. Pin attacks (Appendix A) were emulated by flipping a uniformly random number of bits in at the primary inputs at a uniformly random time. The test setup used and a sample result is shown in Fig. 18c. The error signal was monitored for detected attacks and we observed a 99.998% detection rate for logic attacks (Table VI) at 99.6% for pin attacks.

The 0.002% of undetected logic attacks were those that degraded the compression ratio but did not alter the decompressed word [23]. For example, when the LZ compressor is stuck at an "unmatched" state, meaning no input character ever matches anything in the dictionary, no compression is achieved ( $C_l = 0$ ). Thus, the LZ de-compressor will output the original string instead of reading from its dictionary.

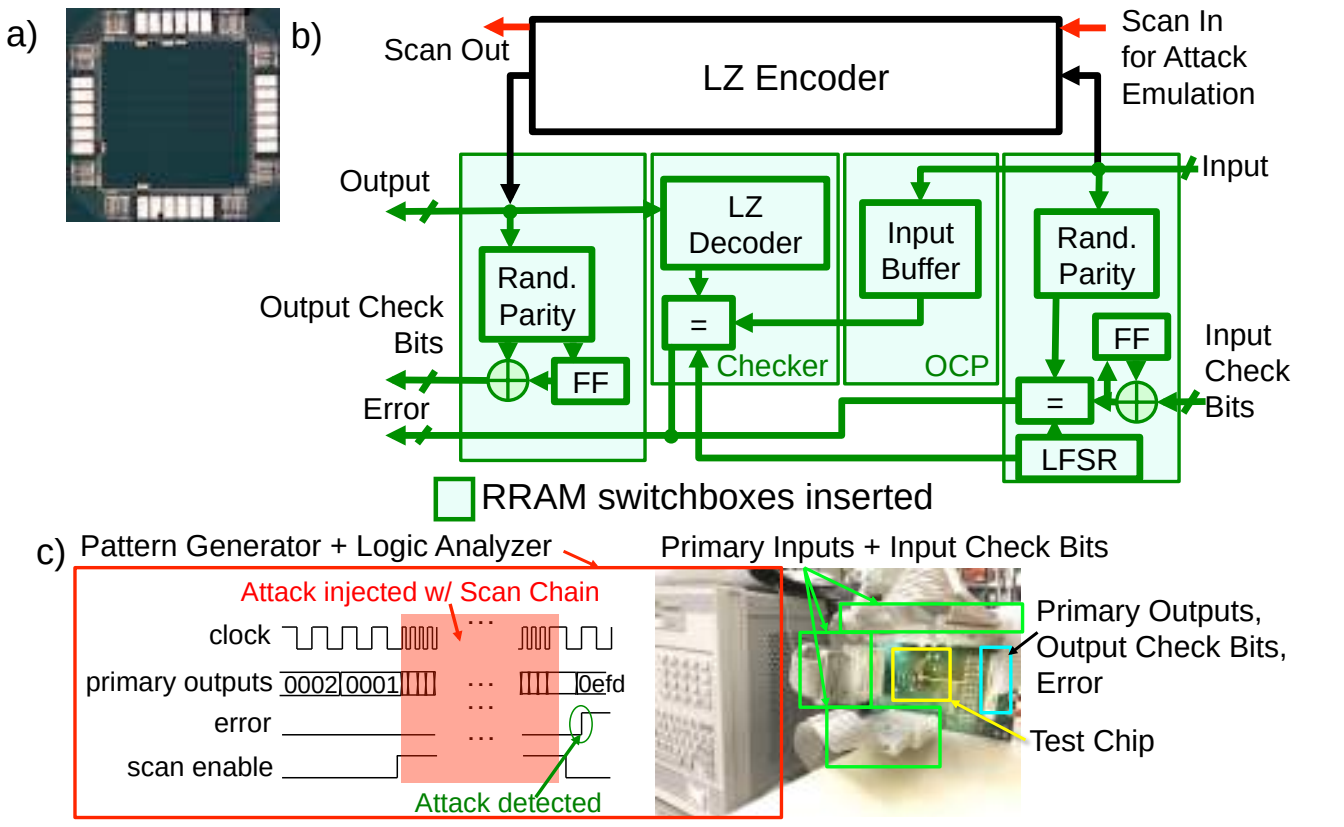

Fig. 18. LZ Encoder design using TPAD. a) Die photo of LZ encoder using TPAD fabricated in 65nm technology. b) Block diagram of an LZ encoder using TPAD. The OCP is an input buffer while the checker is made from the inverse function and an equality checker. 64 RRAM switchboxes were inserted using Algorithm 1 along the cones ending at each of the 24 output bits. c) Test setup for Trojan emulation with sample output result.

The tradeoff between the CR and the area overhead required to secure the design was further analyzed. Configurations of the LZ77 compressor for dictionary sizes ranging from 16 to 256 words, maximum matched length ranging from 4 to 256 words, and input word width ranging from 8 to 32 bits were synthesized and placed-and-routed using the Nangate 45nm standard cell library [40]. The tradeoff between the CR and the area is shown in Fig. 19. The CR is calculated experimentally by compressing a text file containing Shakespeare's Hamlet. As dictionary sizes and input word widths increase, the de-compressor area to compressor area ratio decreases since the de-compressor dictionary is implemented with an SRAM while the compressor dictionary is implemented with a flip-flop-based shift register. Thus, the minimum relative areas were achieved with the maximum dictionary size and maximum input word widths. Increasing the maximum matched length increases the CR without significantly affecting the area overhead of TPAD since only a few gates are changed. However, the CR increase levels off at a maximum matched length of 16 words since it is difficult to find matching substrings beyond 16 characters in Hamlet.

For example, suppose a design requires a CR of 1.38, the area overhead of implementing this design with TPAD using SB programmability would be 34% with a 10% power overhead (shown in Fig. 19a). If SB programmability is not needed, the area overhead would decrease to 32%. Sacrificing a small amount of CR can reduce the area cost further. If 0% area overhead is needed, one can incur a 7% decrease in CR from 1.38 to secure the design (Fig. 19b). The area overhead of SB programmability is 0.04% in this case.

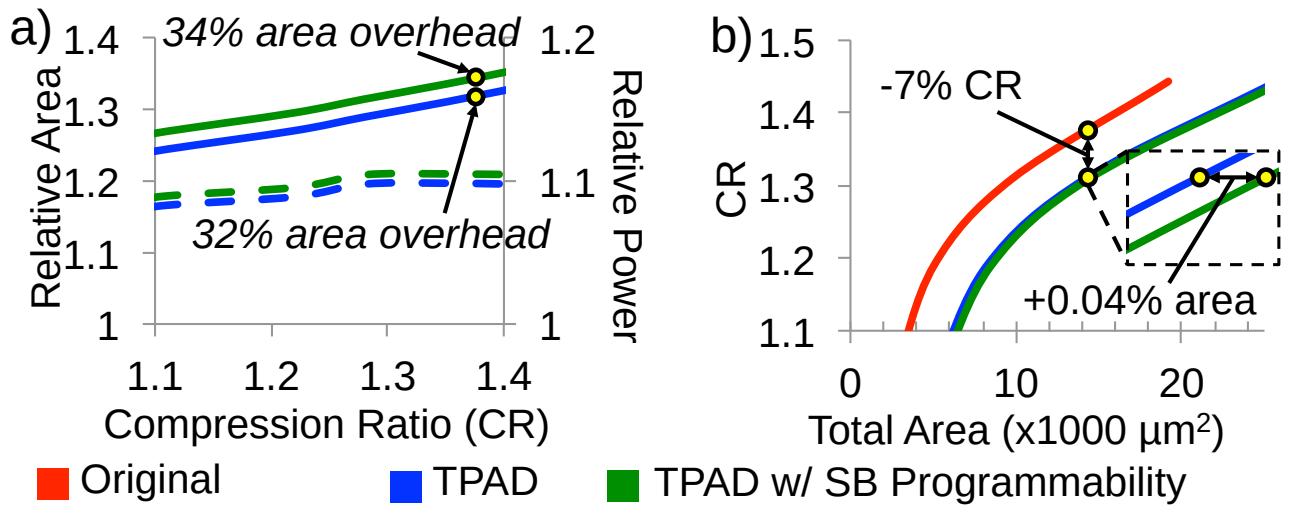

Fig. 19. CR and area tradeoff for LZ encoder design. a) Relative area and power compared to an LZ encoder design without TPAD for fixed CR. b) CR for fixed area. Results shown are after physical design using the Nangate 45nm library.

### *B. Fast Fourier Transform engine*

To further demonstrate the impact that OCP selection can have on the area overhead of TPAD, we implement a pipelined Fast Fourier Transform (FFT) engine. The FFT engine is implemented using the Cooley-Tukey algorithm [41] with half-precision floating-point arithmetic.

While Parseval's theorem is a popular CED method for FFT circuits [36], an adversary can change the output of the FFT such that the sum of squares is still preserved. For example, an adversary can permute the outputs in arbitrary ways and still preserve the sum of squares. Thus, Parseval's theorem cannot be used for Trojan detection. Instead, we make use of the Plancherel theorem for the Discrete Fourier Transform, stated below.

**Plancherel Theorem.** Let $x, y \in \mathbb{C}^N$ have Discrete Fourier Transforms $X, Y \in \mathbb{C}^N$, respectively. Let $y^*, Y^*$ be the complex conjugates of $y, Y$ respectively. Then

$$\sum_{i=0}^{N-1} x_i y_i^* = \frac{1}{N}\sum_{k=0}^{N-1} X_k Y_k^*.$$

To secure an FFT chip, one might think to randomly select a pre-computed Fourier Transform pair $y$, $Y$ and verify that equality in the Plancherel Theorem holds for the observed input-output pair $x, X$. However, since the FFT engine (like many arithmetic circuits in general [44]) is implemented with floating-point arithmetic, roundoff errors could corrupt the output, leading to many false positive Trojan detection outcomes. Hence, CED techniques that use comparisons of floating-point numbers to detect Trojans must have a way of distinguishing between errors due to finite-precision effects and errors due to attacks.

While roundoff has traditionally been analyzed via a probabilistic model [42], it is not at all random (the magnitude of numerical error at the output of a finite-precision computation is a deterministic function of the inputs [43]). TPAD for FFT circuits allows the designer to specify an acceptable level $T$ for numerical errors (which can be chosen based on the specifics of the arithmetic implementation). Errors that exceed this threshold are considered attacks. Explicitly, let $x$ be the input to an $N$-point FFT circuit and let $T$ be the threshold. If the FFT output is $X'$ where

$$\left|\sum_{i=0}^{N-1} x_i y_i^* - \frac{1}{N}\sum_{k=0}^{N-1} X'_k Y_k^*\right| > T$$

then the checker reports a Trojan attack. In hardware, vectors $y$ and $Y$ can be any pre-computed Fourier Transform pair and should be programmed upon startup of the chip. Similar CED methods were analyzed for general floating-point operations in [44]. It was shown that $y^*, Y^*$ could be chosen as periodic vectors with $\|y^*\|_2 = \frac{1}{N}\|Y^*\|_2 = 1$ in order to minimize roundoff while maintaining sensitivity to errors. While an attacker can force $y^*, Y^*$ to zero to bypass the CED test, this requires substantial hardware to be added to the chip and it can easily be detected after manufacturing (Appendix A).

A block diagram of the architecture is shown in Fig. 20a. The OCP computes the inner product of the input and $y^*$ while the checker computes the inner product of the output and $Y^*$ and compares it to the output of the OCP. The design was synthesized and placed-and-routed in Nangate 45nm technology [40]. Fig. 20b shows the overhead compared to the number of points in the FFT. For larger FFT designs, the overhead decreases. The CED circuits take 4% area overhead for a 128-point FFT. I/O encoding to protect against pin attacks incurs another 3.4% area overhead, yielding a total of 7.4% area overhead. The power overheads were found to be the same as the area overheads.

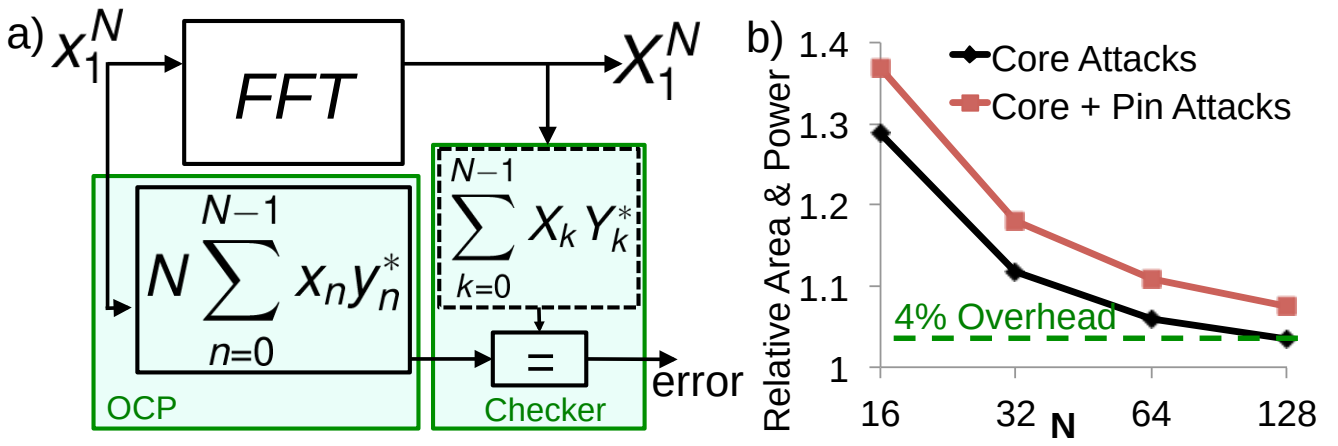

Fig. 20. a) Block diagram of the specialized OCP function implementing the Plancherel theorem. b) Relative area and power compared to those of the FFT.

Logic attack coverage was found to be 99% (Table VI), where the 1% corresponded to Trojans that altered both the OCP and FFT outputs such that the Plancherel theorem was still satisfied for the $y$ and $Y$ vectors chosen. In these simulations, random white noise was used for $y$ and its corresponding FFT was used as $Y$. The vectors used had small norm to prevent overflow and none of the values in $y$ or $Y$ were zero. All electrical and reliability attacks were caught while pin attack coverage remained at 99.6% with 8 check bits.

TABLE VI: COVERAGE AND OVERHEAD FOR SPECIALIZED OCP EXAMPLES

| Area/Power Overhead (%) | | Attack Coverage (%) | | | | |
|---|---|---|---|---|---|---|
| Circuit: | Secure Arch | Random Error | Pin | Logic | Electrical | Reliability |
| LZ77 | 27/9 | 99.998 | 99.6 | 99.998 | 100 | 100 |
| FFT 128 | 7.4/7 | 100 | 99.6 | 99 | 100 | 100 |

## VI. PREVENTION OF CAD TOOL ATTACKS

While switchbox programmability protects against reverse-engineering during fabrication, the design phase is still vulnerable to attacks by CAD tools [6]. Since CAD tools have complete information of the entire chip design, it is possible for them to strategically insert Trojans such that the Trojan detection circuitry cannot detect them (e.g., a Trojan is inserted in both the main function and in the OCP such that predicted check bits and actual check bits are still equal). Verification-based methods for detecting Trojans can be used [14] [104] but are prone to false-negatives (i.e., the Trojan may not be detected). It is likely that all fabricated chips will contain such Trojans; thus, these Trojans can be detected by destructively testing a few chips after fabrication (Section VII. A). However, if CAD tool providers have close relationships with foundries, it may be possible to fabricate a limited amount of chips containing Trojans along with the Trojan-free chips (Section VII. A). A split-design flow with TPAD can ensure Trojan detection (after fabrication).

The split design flow (shown in Fig. 21a) is used to decouple checker information from original design information. As mentioned in Section II, we assume the original RTL is Trojan-free; thus, it is used by a trusted script to generate the RTL for the Trojan detection circuitry (i.e., input decoding, output encoding, logic CED, and error encoding in Fig. 5). The original RTL and Trojan detection RTL are synthesized separately to prevent any CAD tool from receiving complete design information at any time. There must be no retiming during synthesis of each part to preserve the timing for the TPAD architecture in the final netlist. A malicious synthesis or physical design tool would have no knowledge of the other part of the chip when inserting a Trojan.

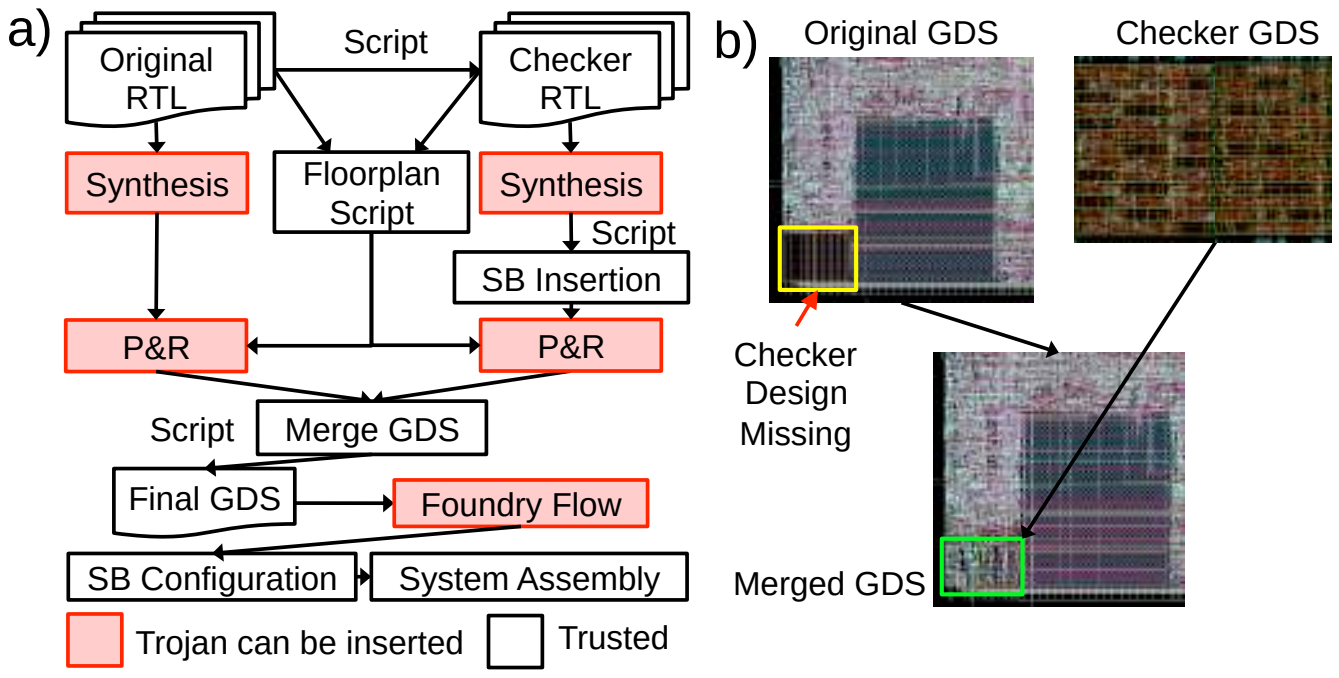

Fig. 21. a) Split design flow for security against compromised CAD tools. The potentially compromised tools are highlighted in red. Synthesis tools and Place and Route tools are potentially compromised. b) Original GDS and checker GDS are merged into a single GDS file.

Fig. 21b illustrates the outcomes of this split-design flow. A trusted floorplanning script is used to partition the die into separate blocks (one for original design and one for Trojan detection circuitry). The original and Trojan detection portions are place-and-routed separately, each keeping its counterpart black-boxed. After physical design, the two layouts are merged with a simple, trusted, script and sent for fabrication. The split design flow is required when malicious CAD tools are assumed in the threat model, regardless of whether or not split-manufacturing is used. Requirements are summarized in Table VII.

Designs were synthesized and place-and-routed with our split-design flow and compared with those without the split-design flow. Specifically, circuits were compared for functionality (LVS), critical path delay, and total area. Our results (Table VIII) show that there is no decrease in performance in the split-design flow compared to a traditional, non-split-design flow.

TABLE VIII: COMPARISON OF SPLIT DESIGN FLOW VS NON-SPLIT FLOW

| Circuit | Performance Loss | Area Difference | LVS Pass |
|---|---|---|---|
| LZ77 | No | 0% | Yes |
| OpenRISC | No | 0% | Yes |
| FFT64 | No | 0% | Yes |

## VII. SURVEY OF TROJAN DETECTION METHODS

### A. Destructive vs. Nondestructive Methods

Methods for detecting hardware Trojans can be divided into destructive and nondestructive methods. *Destructive methods* include post-manufacture tests and/or inspections that render the chip to be non-operational (or change its lifetime reliability characteristics).

Destructive methods are ideal for identifying Trojans that affect all (or a large number of) chips in a batch, since testing a few chips would be enough. For example, Scanning Electron Microscopy (SEM) or Focused Ion Beam (FIB) [52] could be used to destructively delayer chips, and verify transistors and interconnects. Destructive stress tests may be used to detect reliability attacks (Appendix A) [48].

However, this approach becomes difficult when Trojans are only inserted into a select few chips. For example, assume that if a chip containing a Trojan is destructively tested or inspected, the Trojan will be found and that particular chip will be destroyed. Assume that if a Trojan is found, the testing procedure stops and the rest of the batch is declared untrustworthy. Let $N$ be the total number of chips in the batch and let $a$ be the total number of attacked chips. Then, the probability of Trojan detection (depicted in Fig. 22 for various scenarios) by sampling $t$ chips and testing them individually is

$$\mathbb{P}(\text{attack detected}) = 1 - \prod_{i=0}^{t-1}\left(1 - \frac{a}{N-i}\right)$$

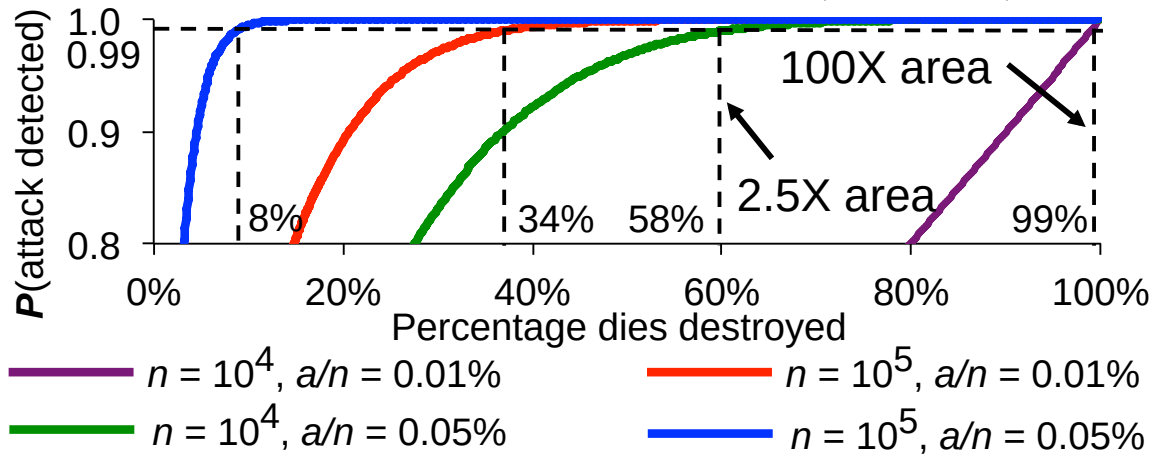

Fig. 22. Probability of detecting a Trojan in a batch of chips with destructive testing of individually sampled chips. Curves are labeled with n: total number of chips in batch and a: number of chips with Trojans.

For a batch of 100000 chips, assuming 0.05% of them contains a Trojan (a=50), 8% of the chips must be destroyed to achieve 99% Trojan detection probability. However, for a batch of only 10000 chips with 0.05% (a=5), of them

TABLE VII: SUMMARY OF HARDWARE REQUIREMENTS UNDER VARIOUS MANUFACTURING AND ATTACK MODELS

| | Manufacturing Technology | | | | | | | | | | | |
|---|---|---|---|---|---|---|---|---|---|---|---|---|
| | No Split-manufacturing | | | | | | Split-manufacturing after M1 | | | | | |
| | CAD Attacks Only | | Foundry Attacks Only | | | Both | CAD Attacks Only | | Foundry Attacks Only | | | Both |
| Feature: | Logic/ Pin/ Electrical | Reverse Engineer | Logic/ Pin/ Electrical | Reverse Engineer | Reliability | All | Logic/ Pin/ Electrical | Reverse Engineer | Logic/ Pin/ Electrical | Reverse Engineer | Reliability | All |
| TPAD Arch. | R | N | R | R | R | R | R | N | R | N | R | R |
| Programmability | R | R | R | R | N | R | N | N | N | N | N | N |
| Split Design | R | R | N | N | N | R | R | R | N | N | N | R |

R – This feature is **required** to protect against the above set of attacks when the above manufacturing technology is used.
N – This feature is **not required** to protect against the above set of attacks when the above manufacturing technology is used.

containing a Trojan, 58% of the chips must be destroyed to achieve the same confidence. If only 0.01% of the chips contained a Trojan (a=1), 99% of the chips would have to be destroyed. Thus, nondestructive methods are needed when a relatively small number of chips are attacked or batch size is small.

## B. *TPAD vs. other techniques*

The following list categorizes various methods for Trojan prevention and detection. A comparison is given in Table IX.

1. Split-Manufacturing: A split at the first metal layer (or very low metal layers) can prevent an untrusted foundry from reverse-engineering a design since most interconnects are missing [46][47]. However, split manufacturing does not provide Trojan detection by itself. Destructive stress tests on selected back-end-of-line stacks can detect reliability attacks [48]. Trusted monitor chips stacked on top of an untrusted chip using through-silicon vias (TSVs) may be used to actively detect attacks [49] or selected wires may be lifted (using TSVs) to a secure layer to obfuscate a design [106]. Large TSV pitch can lead to area inefficiencies [50].
2. Imaging: Trojans can be detected by Electromagnetic radiation (EMR) imaging to search for physical anomalies [51][53]. This approach does not prevent an adversary from reverse-engineering the design. Scanning Electron Microscopy (SEM) or Focused Ion Beam (FIB) can also be used, but they require delayering the chip, thus destroying it [52].
3. Logic Encryption: These nondestructive techniques prevent Trojans by placing additional logic gates and making them partially controllable [54][55]. These methods do not offer Trojan detection. [56] is a concurrent method to prevent Trojans by using external modules to control operation. However, delay penalties are introduced and additional pins are required.
4. FSM Analysis: The state transition diagram of a circuit is modified so that a specific sequence of transitions must take place before the circuit operates according to the initial design [57][58]. Since an untrusted foundry could have access to the entire design, the state transitions can possibly be inferred. RTL-level Trojans may be detected by identifying signals that rarely change [59][60]. This approach can result in false positives and it cannot detect any Trojans inserted by the foundry.
5. Active Monitoring: These nondestructive techniques allow for Trojan detection by monitoring (some of) a chip's functional behavior during runtime [61]-[66] (e.g. control signals in a processor). [61] is limited to RTL-level Trojans. [62] requires a snapshot of all I/O signals of a chip to be obtained periodically and analyzed for attacks. In addition, these techniques only monitor a subset of signals (and the properties to be checked may not be exhaustive), leaving the systems vulnerable to attacks. [63]-[66] require reconfigurable logic (e.g., FPGAs), thus greatly increasing area overheads.
6. Delay Fingerprinting: These non-destructive techniques detect Trojans by taking delay measurements along certain paths and finding a difference between them and timing for an intended design [67]-[69]. If a Trojan is inserted, the delay may not match the model. While [70][71] present strategies for inserting test points to make various paths observable, these methods are prone to false positives since Trojans must be distinguished from variability and noise [89].
7. Switching Power Analysis: These techniques detect Trojans by measuring dynamic current waveforms and comparing with a model for an intended design [74]-[79]. If there is a significant difference between the two current waveforms, a Trojan is assumed. [78][79] are concurrent solutions that perform classification in real-time. These methods are prone to false positives since Trojans must be distinguished from variability and noise.
8. Gate Level Characterization (GLC): Measurements are taken, leakage power [81]-[83] or delays of individual gates

TABLE IX: COMPARISON OF DIFFERENT TROJAN DETECTION AND PREVENTION SCHEMES

| Approach | Refs | Prevention or Detection | Destructive | Split-manufacture Required? | Attacks Covered | Area Overhead | Concurrent | False Positives |
|---|---|---|---|---|---|---|---|---|
| TPAD | This paper | Both | No | No | 1-6 | 7.4 - 165% based on design | Yes | No |
| Split after M1 | [46]-[47] | Prevention | No | Yes | 1 | 0 % | No | N/A |
| Test-only BEOL | [48] | Both | Yes | Yes | 1-5 (~) | Dies destroyed | No | No |
| 3D chip stacking | [49] | Both | No | Yes | 1, 3-6 | TSVs, Control Chip | Yes | No |
| EMR Imaging | [51],[53] | Detection | No | No | 2-4 (^+) | 0 % | No | No |
| SEM or FIB | [52] | Detection | Yes | No | 2-5 (~) | Dies destroyed | No | No |
| XOR or LUT Insertion | [54][55] | Prevention | No | No | 1 | Gates + Pins | No | N/A |
| Silencing | [56] | Prevention | No | No | 1-3($) | Encryption, Input Reorder | Yes | N/A |
| HARPOON | [57][58] | Prevention | No | No | 1 | 5 – 20 % | No | N/A |
| FANCI, FIGHT | [59][60] | Prevention | No | No | 1 ($) | 0 % | No | Yes |
| Control Monitoring | [61][62] | Detection | No | Yes (#) | 2-6(&) | Monitoring Circuits | Yes | No |
| FPGA w/ Checking | [63]-[66] | Both | No | No | 1, 3-6 | FPGA | Yes | No |
| On-chip sensing of delays | [67]-[71] | Detection | No | Yes (#) | 2-6(&*) | Second clock, latches, sensors. | Can be | Yes |
| BISA | [72][73] | Detection | No | No | 2-4(&*) | ATPG logic and pins | No | No |
| TeSR | [74][75] | Detection | No | Yes (#) | 2-4(&*) | Retry+Sense+Compare logic | No | Yes |
| Dynamic power | [76][77] | Detection | No | No | 2-4 (*^) | Dummy FFs | No | Yes |
| GLC | [81]-[84] | Detection | No | No | 2-4 (*^) | 0 % | No | Yes |
| Multimodal Fingerprinting | [85]-[88] | Detection | No | Can be (#) | 2-4 (*^) | Can be 0% | Can be | Yes |
| Random Test Patterns | [90]-[91] | Detection | No | No | 2-4 (^) | 0 % | No | No |
| ODETTE/ VITAMIN | [92]-[93] | Both | No | Yes (#) | 1-4 (^) | $\boldsymbol{Q}$ and $\bar{\boldsymbol{Q}}$ pins | No | No |

Attack Coverage Meanings: 1-Prevents Reverse-Engineering, 2-Detects Pin Attacks, 3-Detects Logic Attacks, 4-Detects Electrical Attacks, 5-Detects Reliability Attacks, 6-Detects Circuit Errors. Conditions Meanings: *-Determined by variability/confidence interval, &-Only along monitored paths, ^- Only those found by test vectors, +-Only top/bottom layers and uncongested middle, ~-Remaining chips insecure, $-RTL Level Only, #-Security monitors built by trusted foundry

[84] are estimated, and Trojans are identified. GLC shares the limitations of other fingerprinting techniques (e.g., [89]).

9. Multimode Fingerprinting: Several fingerprinting methods are combined and GLC can be used to compare the results with a model [85]-[88]. While stronger, these methods share similar limitations as GLC.
10. Activation by Test Vectors: Random test patterns are used to attempt to activate Trojans during post-manufacture testing [90]-[93]. [92]-[93] introduce programmability to accelerate Trojan activation during post-manufacture testing and to provide logic encryption. However, these methods are non-concurrent; thus, they are not able to detect Trojans that activate after post-manufacture testing (e.g. time-bombs, reliability attacks, etc.).

## VIII. Conclusions

TPAD protects digital systems from hardware Trojan attacks through a combination of special concurrent error detection techniques and selective programmability. As a result, it can prevent and detect hardware Trojans with high probability and with no false positives. We demonstrated a variety of detection techniques, from general randomized parity coding (that can be expensive for general designs) to specialized detection techniques (tailored for special functions) that significantly reduce TPAD area overheads. Hardware test chip results demonstrate the effectiveness and practicality of TPAD.

While many aspects of TPAD have been extensively explored in this paper, several opportunities for future work remain. These include:

1. SB insertion algorithms that avoid degenerate cases and/or ensure that multiple checking circuits can be created (Section III).

2. Software checking techniques that can complement TPAD for systems supporting both hardware and software programmability.

3. Techniques for recovering from hardware Trojan attacks (Appendix B).

## Appendix A: Types of Hardware Trojans

We categorize hardware Trojans by their physical locality. We split core attacks into subcategories relating to the type of circuit element modified and decoupling attacks into subcategories relating to the specific details of the attack.

1. **Pin Attacks**: The logical value of a pin is flipped (Illustrated in Fig. 23a) before the core logic of the chip.
2. **Core Attacks**: The core logic is modified such that at least one incorrect output sequence is produced during operation (Illustrated in Fig. 23b). This includes:
   a. *Logic attacks*: additional circuitry is added or existing circuitry is modified to cause erroneous outputs of a logic block (Fig. 1a).
   b. *Reliability attacks*: Modification to transistors or wires to cause physics-induced failures over time (Fig. 1c).
   c. *Electrical attacks*: Modification to the wiring of the chip to change the timing of the circuit (Fig. 1b).
3. **Decoupling attacks:** The TPAD checking circuitry is bypassed and fed a set of valid inputs while incorrect data is sent through the main logic (Fig. 23c). This includes:
   a. *Stored state:* A set of flip-flops stores an earlier (valid) state of the system. At some arbitrary point in time, this state is sent through the logic CED (thereby forcing valid error signals), while arbitrary changes are made to the main logic (Fig. 23c). If $n$ is the number of flip-flops in the design, this attack requires at least $n$ extra flip-flops.
   b. *FFT zeroing:* When the Plancherel Theorem is used as a CED technique for FFT circuits (Section V.B), one can pull $y^*$ and $Y^*$ to zero to force a valid error signal. Permanent grounding is very simple to detect, as one can always choose a $y$ and $Y$ that are not corresponding FFT pairs as part of a non-destructive post-manufacture test and check to that the error monitor actually reports an error. The more stealthy case for such an attack would be to add a single pull-down transistor and a pass-transistor for each of the non-sign bits of the $y^*$ and $Y^*$ signals. Thus, it would require at least $120N$ additional minimum-sized transistors to attack an $N$-point half-precision floating-point FFT (15 non-sign bits in a half-precision floating point word; 2 transistors needed per bit; $y^*$,$Y^*$ each have a real word and imaginary word).
   c. *Randomized parity nulling:* Because randomized parity codes are linear (Section IV.A), the all-zero sequence is always a codeword [24]. If randomized parity codes are used in output encoding (Section II.A), the primary outputs can be changed to all zeros and the previous output check bits can be reused. This would evade the input decoding procedure (Section II.B). This attack can be prevented if a nonlinear code is used instead of a linear code [35]. This attack requires at least 2 transistors per primary output bit and 1 flip-flop per output check bit.

For the circuits analyzed in this work, Table X gives a lower bound (i.e., minimum of attacks 3a-3c) on the additional area and transistors needed to implement a decoupling attack in different technology nodes, when the extra hardware is *added* to a 2D chip. These attacks require enough hardware, that they can be detected by the following sequence of non-destructive post-manufacture inspections:

1. Un-packaged chips: the backside of every chip can be imaged using an X-ray imaging technique [53], revealing the transistor contacts (if their dimensions are > 30nm). Total number of transistors can then be estimated and compared with the GDS. If contact dimensions are < 30nm, then optical or IR imaging, with 2-3$\mu$m resolution, may be used to compare the total area with the GDS [80] [94].
2. Packaged chips: the package of every chip must be removed and the backside of the chips must be polished non-destructively [94]. Then inspect them as un-packaged chips.

The chips must be packaged or re-packaged in a trusted facility before being used in the field.

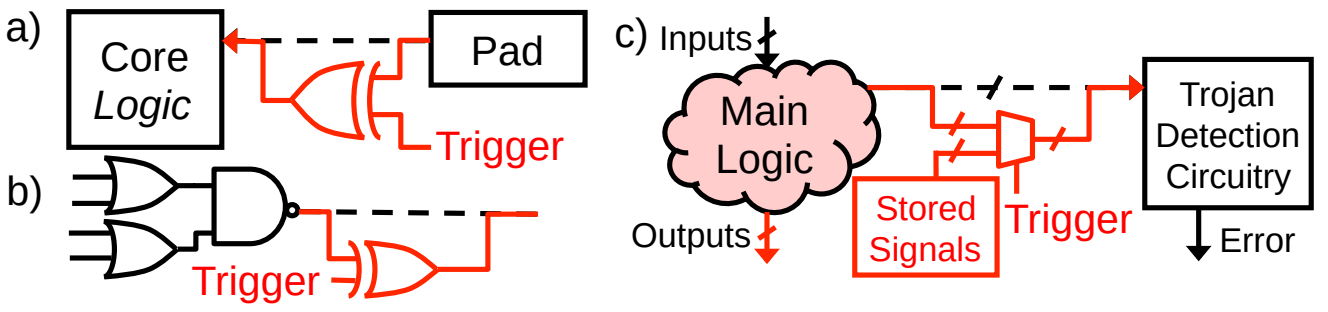

Fig. 23. Types of Trojans. a) Pin attack: the inputs to the chip are altered before any computation by the core logic. b) Core attack: the Boolean output of a function inside the core is changed. c) Decoupling attack: the Trojan detection circuitry is fed with correct stimuli while the main logic is hijacked.

While the above steps are sufficient if the decoupling attack hardware is *added* to a chip, they may not be sufficient if the adversary attempts to *replace existing filler cells* [107] within a design with the decoupling attack hardware. If there are a large number of filler cells within a design, an adversary at the foundry can remove these filler cells and fit the circuitry for an attack within the resulting empty area. With no net area increase, and if the contacts are sufficiently small (<30nm), attacks are difficult to detect with standard X-Ray, IR, or optical imaging. To prevent such scenarios, [107] provides a strategy for designing custom filler cells with a very high reflectance at near-IR wavelengths. Attacks that replace these filler cells can then be detected using backside optical imaging [107].

TABLE X. MINIMUM ADDITIONAL OVERHEAD FOR DECOUPLING ATTACK

| Circuit: | Area ($\mu m^2$) | | | Number of Transistors | Attack |
|---|---|---|---|---|---|
| | 45nm | 28nm | 15nm | | |
| OpenRISC (*r* = 8) | 39 | 18 | 10.7 | 396 | *3c* |
| LZ77 (*r* = 8) | 36 | 17 | 10.3 | 214 | *3c* |
| FFT (128-point) | 230 | 83 | 37 | 15360 | *3b* |

Additionally, *denial of service (DoS)* attacks may be attempted during chip design and fabrication. DoS attacks during design time include wrong synthesis, ignored design constraints, physical design tools refusing to place and/or route correctly, and tools reporting wrong values for timing. By using a split design flow (Section VI), tools do not know the entire design at the time of synthesis and layout. Thus, if design-time DoS attacks are not caught during verification, they cannot be made in a way that evades detection by the checking circuitry during runtime. If DoS attacks are caught during verification, re-synthesis or re-P&R with another tool may allow for recovery. However, recovery in the field can be difficult since the wrong chip was manufactured (Appendix B). Another family of DoS attacks can happen at the foundry. These include the foundry skipping certain metal layers or using technology parameters other than the ones specified to the designer. These types of attacks may be caught in post-manufacture testing. Recovery, however, is not an option.

The last family of DoS attacks only manifest after fabrication. They include chips (or some parts of a chip) turning off after a rare logic sequence or at a specific time. TPAD concurrently detects such attacks, since the parts of the chip that fail to turn on will produce incorrect error signals. However, these attacks can be difficult to recover from as they could cause permanent shorts or open circuits in the chip.

## APPENDIX B: OVERVIEW OF RECOVERY TECHNIQUES

Trojan detection can trigger on-line recovery and repair modes. Potential recovery methods differ depending on the type of Trojan. For instance, methods based on time redundancy such as detection and retry [95], checkpointing and rollback [96], and CRTR and SRTR [97] in multi-threaded parallel systems can be tried to recover the chip. However, these techniques only work when Trojans are *transient*, meaning they are triggered by time and not by a specific state of the chip, and they do not cause permanent damage such as shorting the power signals. For Trojans that are not necessarily transient, error-correction [24] using codes stronger than those of Section IV (requires additional error correction hardware) could be used for recovery. In distributed systems with hardware redundancy, if only a small proportion of chips are infected with a permanent or state-triggered Trojan, Spares [19], N-Modular Redundancy [98], Consensus [99], Coordinated Checkpointing [100], or Byzantine Fault Tolerance in distributed systems [101] can be implemented to recover from attacks at the cost of additional area overhead.

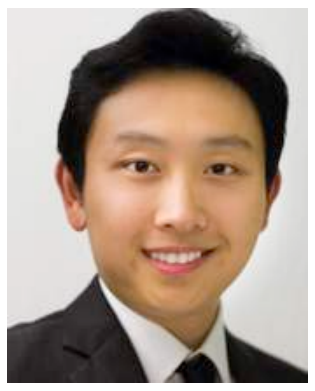

**Tony Wu** received the B.S. degree in EE from California Institute of Technology, Pasadena, CA, USA, in 2011, and the M.S. degree in EE from Stanford University, Stanford, CA, USA, in 2013, where he is currently pursuing the Ph.D. degree. His current research interests include design and fabrication of monolithic 3D integrated systems using emerging technologies.

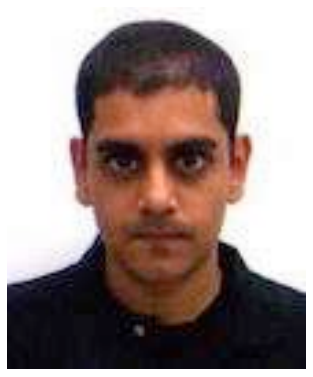

**Karthik Ganesan** received the B.S. degree in EECS and the B.A. degree in Statistics from University of California at Berkeley, Berkeley, CA in 2013, and the M.S. degree in EE from Stanford University, Stanford, CA in 2015, where he is currently pursuing the Ph.D. degree. His research interests include coding theory, applied probability, and their uses in reliable computing and low-power systems design.

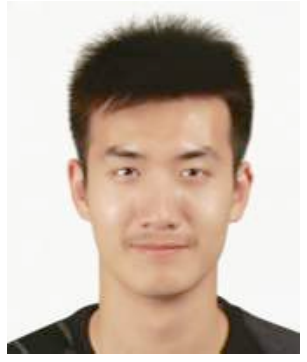

**Yunqing Hu** received the B.S. degree in EE from California Institute of Technology, Pasadena, CA, USA, in 2012, and the M.S. degree in EE from Stanford University, Stanford, CA, USA, in 2014, where he is currently pursuing the Ph.D. degree. His current research interests include design and fabrication of secure 3D integrated systems using emerging technologies.

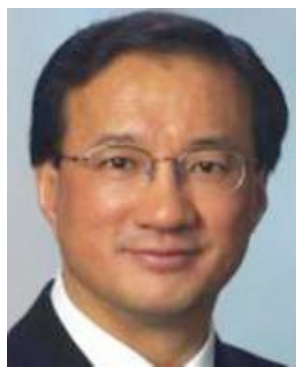

**Simon Wong** received the Bachelor degrees in Electrical Engineering and Mechanical Engineering from the University of Minnesota, and the MS and PhD degrees in Electrical Engineering from the University of California at Berkeley. His industrial experience includes semiconductor memory design at National Semiconductor (78-80) and semiconductor technology development at Hewlett Packard Labs (80-85). He was an Assistant Professor at Cornell University (85-88). Since 1988, he has been with Stanford University where he is now Professor of Electrical Engineering. His current research concentrates on understanding and overcoming the factors that limit performance in devices, interconnections, on-chip components and packages. He is on the board of Pericom Semiconductor.

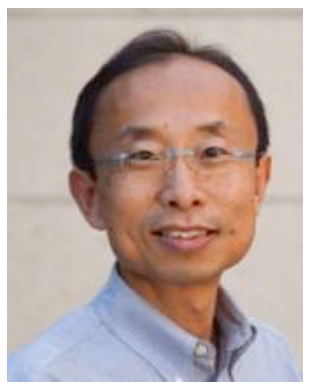

**H.-S. Philip Wong** (F'01) is the Willard R. and Inez Kerr Bell Professor in the School of Engineering. He joined Stanford University as Professor of Electrical Engineering in September, 2004. From 1988 to 2004, he was with the IBM T.J. Watson Research Center. At IBM, he held various positions from Research Staff Member to Manager, and Senior Manager. While he was Senior Manager, he had the responsibility of shaping and executing IBM's strategy on nanoscale science and technology as well as exploratory silicon devices and semiconductor technology. His present research covers a broad range of topics including carbon electronics, 2D layered materials, wireless implantable biosensors, directed self-assembly, nanoelectromechanical relays, device modeling, brain-inspired computing, and non-volatile memory devices such as phase change memory and metal oxide resistance change memory. He is a Fellow of the IEEE and served on the Electron Devices Society AdCom as elected member (2001 – 2006). He served as the Editor-in-Chief of the IEEE Transactions on Nanotechnology in 2005 – 2006, sub-committee Chair of the ISSCC (2003 – 2004), General Chair of the IEDM (2007), and is currently a member of the Executive Committee of the Symposia of VLSI Technology and Circuits (2007 – present). He received the B.Sc. (Hons.), M.S., and Ph.D. from the University of Hong Kong, Stony Brook University, and Lehigh University, respectively. His academic appointments include the Chair of Excellence of the French Nanosciences Foundation, Guest Professor of Peking University, Honorary Professor of the Institute of Microelectronics of the Chinese Academy of Sciences, Visiting Chair Professor of Nanoelectronics of the Hong Kong Polytechnic University, and the Honorary Doctorate degree from the Institut Polytechnique de Grenoble, France. He is the founding Faculty Co-Director of the Stanford SystemX Alliance – an industrial affiliate program focused on building systems.

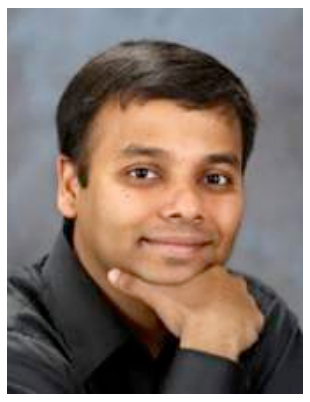

**Subhasish Mitra** (F'13) directs the Robust Systems Group in the Department of Electrical Engineering and the Department of Computer Science of Stanford University, where he is the Chambers Faculty Scholar of Engineering. Before joining Stanford, he was a Principal Engineer at Intel.

Prof. Mitra's research interests include robust systems, VLSI design, CAD, validation and test, emerging nanotechnologies, and emerging neuroscience applications. His X-Compact technique for test compression has been key to cost-effective manufacturing and high-quality testing of a vast majority of electronic systems, including numerous Intel products. X-Compact and its derivatives have been implemented in widely-used commercial Electronic Design Automation tools. His work on carbon nanotube imperfection-immune digital VLSI, jointly with his students and collaborators, resulted in the demonstration of the first carbon nanotube computer, and it was featured on the cover of NATURE. The US National Science Foundation presented this work as a Research Highlight to the US Congress, and it also was highlighted as "an important, scientific breakthrough" by the BBC, Economist, EE Times, IEEE Spectrum, MIT Technology Review, National Public Radio, New York Times, Scientific American, Time, Wall Street Journal, Washington Post, and numerous other organizations worldwide.

Prof. Mitra's honors include the Presidential Early Career Award for Scientists and Engineers from the White House, the highest US honor for early-career outstanding scientists and engineers, ACM SIGDA/IEEE CEDA A. Richard Newton Technical Impact Award in Electronic Design Automation, "a test of time honor" for an outstanding technical contribution, the Semiconductor Research Corporation's Technical Excellence Award, and the Intel Achievement Award, Intel's highest corporate honor. He and his students published several award-winning papers at major venues: IEEE/ACM Design Automation Conference, IEEE International Solid-State Circuits Conference, IEEE International Test Conference, IEEE Transactions on CAD, IEEE VLSI Test Symposium, Intel Design and Test Technology Conference, and the Symposium on VLSI Technology. At Stanford, he has been honored several times by graduating seniors "for being important to them during their time at Stanford."

Prof. Mitra has served on numerous conference committees and journal editorial boards. He served on DARPA's Information Science and Technology Board as an invited member. He is a Fellow of the ACM and the IEEE.